\documentclass[12pt,preprint]{aastex}

\usepackage{graphicx, color}% Include figure files
\usepackage{dcolumn}% Align table columns on decimal point
\usepackage{bm}% bold math
\usepackage{booktabs}
\usepackage{natbib}
\usepackage{lscape}

\newcommand {\vect}[1]{\mbox{\boldmath $#1$}}

\newcommand {\inty}[2]{\int_{#1}^{#2}}

\newcommand {\dif}[3][]{\frac{d^{#1}#2}{d#3^{#1}}}
\newcommand {\pdif}[3][]{\frac{\partial^{#1}#2}{\partial#3^{#1}}}

\newcommand {\lsim}{\hspace{0.3em}\raisebox{0.4ex}{$<$}\hspace{-0.75em}\raisebox{-.7ex}{$\sim$}\hspace{0.3em}}

\makeatletter
\def\mart{\@ifnextchar[{\mart@@}{\mart@}}
\def\mart@@[#1]#2{\sqrt[#1]{\mathstrut{#2}}}
\def\mart@#1{\sqrt{\mathstrut{#1}}}
\makeatother
\newcommand {\Alfven}{Alfv\'{e}n}
\newcommand{\myemail}{minoshim@stelab.nagoya-u.ac.jp}

\newcommand{\RHESSI}{\it RHESSI}

\newcommand{\Hinode}{\it Hinode}

\newcommand{\FP}{Fokker-Planck}

\newcommand{\exb}{{\vect E} \times {\vect B}}
\newcommand{\insitu}{\it in-situ}

\long\def\symbolfootnote[#1]#2{\begingroup%
\def\thefootnote{\fnsymbol{footnote}}\footnote[#1]{#2}\endgroup}

\begin{document}

\title{Drift-Kinetic Modeling of Particle Acceleration and Transport in Solar Flares}

\author{T. Minoshima\footnote{Current address: Institute for Research on Earth Evolution, Japan Agency for Marine-Earth Science and Technology, 3173-25, Syowa-machi, Kanazawaku, Yokohama 236-0001, Japan}, S. Masuda, and Y. Miyoshi}
\affil{Solar-Terrestrial Environment Laboratory, Nagoya University,
Furo-cho, Chikusa-ku, Nagoya 464-8601, Japan;}
% \author{T. Minoshima\altaffilmark{1}, S. Masuda\altaffilmark{1}, and Y. Miyoshi\altaffilmark{1}}
% \altaffiltext{1}{
% Solar-Terrestrial Environment Laboratory, Nagoya University,
% Furo-cho, Chikusa-ku, Nagoya 464-8601, Japan;
% }
\email{\myemail}

\begin{abstract}
Based on the drift-kinetic theory, we develop a model for particle acceleration and transport in solar flares. The model describes the evolution of the particle distribution function by means of a numerical simulation of the drift-kinetic Vlasov equation, which allows us to directly compare simulation results with observations within an actual parameter range of the solar corona.
% Using this model, we investigate the time evolution of the electron distribution in electromagnetic fields of a flare.
Using this model, we investigate the time evolution of the electron distribution in a flaring region.
The simulation identifies two dominant mechanisms of electron acceleration.
 One is the betatron acceleration at the top of closed loops, which enhances the electron velocity perpendicular to the magnetic field line. The other is the inertia drift acceleration in open magnetic field lines, which produces antisunward electrons.
% The resulting velocity space distribution significantly deviates from an isotropic distribution, due to both the acceleration and transport effects. 
The resulting velocity space distribution significantly deviates from an isotropic distribution.
% These two accelerations are in agreement with typical observational features of the flare particle acceleration; loop-top hard X-ray and microwave emissions by energetic electrons, and type III radio bursts by escaping electrons from the Sun that are observed by {\insitu} measurements in interplanetary space.
The former acceleration can be a generation mechanism of electrons that radiate loop-top nonthermal emissions, and the latter be of escaping electrons from the Sun that should be observed by {\insitu} measurements in interplanetary space and resulting radio bursts through plasma instabilities.
\end{abstract}


\section{Introduction}\label{sec:introduction}
{ Many observations with such as hard X-rays (HXRs), $\gamma$-rays, and microwaves have revealed that a solar flare is one of the strongest particle accelerator in our solar system. Electrons that are accelerated to several tens of keV to MeV radiate HXRs at footpoints of flare loops \citep[e.g.,][]{1994PhDT.......335S,2003ApJ...595L.103K,2009ApJ...697..843M}, and sometimes above the top of a soft X-ray loop \citep{1994Natur.371..495M,1995PASJ...47..677M}. They are also observed via microwaves in the gigahertz band. These emissions have been used to understand the properties of accelerated electrons \citep[e.g.,][]{1983Natur.305..292N,1988ApJ...324.1118K,2000ApJ...545.1116S,2001ApJ...557..880K,2002ApJ...576..505W,2008ApJ...673..598M}. Recently, accelerated ions have been studied by the $\gamma$-ray observations of the {\it Reuven Ramaty High Energy Solar Spectroscopic Imager} \citep[{\it RHESSI};][]{2003ApJ...595L..69L,2006ApJ...644L..93H}.}

In addition to these observations of accelerated particles at flare sites, escaping particles from the Sun into interplanetary space, the so-called solar energetic particles (SEPs), and resulting radio bursts are observed in association with flares \citep[e.g.,][]{1982ApJ...253..949L,1985SoPh..100..537L,1998ApJ...503..435E,1999SSRv...90..413R,1999ApJ...519..864K,2007ApJ...663L.109K,2009ApJ...691..806K}. 
% Type III radio bursts \citep[e.g.,][]{1998ARA&A..36..131B} are thought to be due to electromagnetic conversion of Langmuir waves, which are excited through electrostatic instabilities of streaming electrons along open magnetic field lines \citep{2009JGRA..11402104L}. 
% Spacecrafts in interplanetary space have carried out {\insitu} observations of SEPs \citep[e.g.,][]{1982ApJ...253..949L,1985SoPh..100..537L,1998ApJ...503..435E,1999SSRv...90..413R,1999ApJ...519..864K,2007ApJ...663L.109K,2009ApJ...691..806K}. 
\cite{2007ApJ...663L.109K,2009ApJ...691..806K} have reported detailed examinations of the relationship of electrons between at the flare site and in interplanetary space, by using {\RHESSI} and {\it WIND} observations.
% They have confirmed their temporal coincidence, but also found the discrepancy of their energy distributions. This may reflect different acceleration processes and/or an energy-dependent propagation effect of electrons.
On the other hand, escaping electrons are often observed with no corresponding flare activity \citep[e.g.,][]{1980ApJ...236L..97P,2003GeoRL..30m..30G,2007ApJ...657..567M,2010ApJ...708L..95E}. This indicates that small-scale particle accelerations frequently take place in the corona.

% Many authors have proposed a model to explain the particle acceleration in solar flares, e.g., magnetic reconnection \citep{2005JGRA..11010215H,2006Natur.443..553D}, waves \citep{1996ApJ...461..445M,2004ApJ...610..550P}, and shocks \citep{1998ApJ...495L..67T}. These models greatly contribute to the understanding of how particles are accelerated to observed energies. For the complete understanding of many observed features, we should take into account the particle transport and dissipation processes as well as the acceleration in the realistic environment of the solar corona. This is because observed quantities do not necessarily reflect the particle distribution just at the acceleration site, but are the convolution of all of these processes \citep{1998ApJ...502..455A}. 

{ Based on the magnetic reconnection scenario \citep{1995ApJ...451L..83S}, many authors have proposed models to explain the particle acceleration in solar flares. The particle acceleration in and near the reconnection region has been well discussed. For example, \cite{1996ApJ...462..997L} has studied the DC electric field acceleration in reconnecting current sheets. \cite{2001JGR...10625979H} have discussed the nonadiabatic acceleration in the vicinity of the reconnection region at which the gyro radius is comparable to the curvature radius of the magnetic field line. Fermi acceleration by the contraction of magnetic islands has been studied by \cite{2006Natur.443..553D}.}
{ In addition to the acceleration near the reconnection region, stochastic acceleration mechanisms in the reconnection downstream region have been proposed by e.g., \cite{1996ApJ...461..445M} (fast-mode magnetohydrodynamic waves) and \cite{1998ApJ...495L..67T} (oblique shocks), although necessary high-frequency waves for the particle scattering is poorly known in the solar circumstance.}

{The models greatly contribute to the understanding of how particles are accelerated to observed energies. For the complete understanding of many observed features, however, we should take into account the particle transport and dissipation processes as well as the acceleration in a realistic environment of the solar corona. This is because observed quantities do not necessarily reflect the particle distribution just at the acceleration site, but are the convolution of all of these processes \citep{1998ApJ...502..455A}}.

% \cite{1997ApJ...485..859S} have proposed a scenario called ``collapsing trap'', in which the particle acceleration, transport, and dissipation in the shrinking magnetic loop are treated simultaneously.
\cite{1997ApJ...485..859S} have proposed a model called ``collapsing trap'', in which the particle acceleration, transport, and dissipation naturally follow from the shrinkage of magnetic loops driven by reconnection.
 % The shrinkage of the loop is thought to be caused by magnetic reconnection \citep[e.g.,][]{1995ApJ...451L..83S}. 
% The scenario is very similar to the ``dipolarization'' in the Earth's magnetosphere.
% Since the particle gyro scale ($\sim 10^{-9} \; {\rm s} \; {\rm and} \sim 10^{0} \; {\rm cm}$ for electrons) is much smaller than the flare scale ($\sim 10^{2} \; {\rm s} \; {\rm and} \sim 10^{9} \; {\rm cm}$) due to very strong magnetic fields, the scenario reasonably assumes the adiabaticity of particles (the guiding-center approximation).
% With this approximation, the scenario can describe a very wide region of flares, except the magnetic dissipation region in which the guiding-center approximation breaks.
These processes are almost adiabatic in the trap, because the particle gyro scale ($\sim 10^{-9} \; {\rm s} \; {\rm and} \sim 10^{0} \; {\rm cm}$ for electrons) is much smaller than the flare scale ($\sim 10^{2} \; {\rm s} \; {\rm and} \sim 10^{9} \; {\rm cm}$) due to very strong magnetic fields in the corona.
The adiabatic acceleration may take place everywhere in the trap, which diminishes a serious problem that the number of accelerated particles required for observed HXR intensities is comparable to the number in a whole flare region \citep{1974IAUS...57..105K,1997JGR...10214631M}.
{ The mechanism can be regarded as the subsequent acceleration process that occurs after the acceleration near the reconnection region, and as the co-existing process with the acceleration in the downstream region.}
% The adiabatic acceleration may take place everywhere in the trap, which diminishes a serious problem that a significant fraction of energies stored in a flare region is expended for the acceleration of particles \citep{1982SoPh...81..137D,1997JGR...10214631M}.
It is noteworthy to note that the model is very similar to the ``dipolarization'' associated with substorms in the terrestrial magnetosphere.

% \cite{1997ApJ...485..859S} have proposed a scenario called ``collapsing trap'', in which particle acceleration, transport, and dissipation in the shrinking magnetic loop associated with flares are considered simultaneously, by assuming the adiabaticity of particles (the guiding-center approximation). 
% This scenario is very similar to the ``dipolarization'' in the Earth's magnetosphere. 
% The scenario assumes the adiabaticity of particles (the guiding-center approximation). 
% This assumption looks reasonable especially in solar flares, because the gyro scale ($\sim 10^{-9} \; {\rm s} \; {\rm and} \sim 10^{0} \; {\rm cm}$ for electrons) is much smaller than the flare scale ($\sim 10^{2} \; {\rm s} \; {\rm and} \sim 10^{9} \; {\rm cm}$). Therefore the scenario can cover very wide region of flares, except the magnetic dissipation region in which the guiding-center approximation breaks.

In the collapsing trap, particles can adiabatically gain energy from convection electric fields ${\vect E} = -{\vect v} \times {\vect B}$ through $\nabla B$ and inertia drift motions.
\cite{2004A&A...419.1159K} have studied the electron distribution function in the shrinking loop, with considering only the acceleration due to the $\nabla B$ drift (the betatron acceleration). 
\cite{2004ApJ...608..554A} has analytically considered the electron transport and the resulting impulsive HXR emissions in the loop.
% \cite{2004ApJ...608..554A} has modeled the electron transport in the shrinking loop from a cusp to a force-free configuration, to explain observed HXR light curves.
 \cite{2005ApJ...635..636G} have presented a rigorous treatment of the adiabatic motion of a single particle in the loop, by numerically solving guiding-center equations of motion. 
\cite{2006ApJ...647.1472K} have performed a guiding-center test particle simulation in a cusp-shaped loop obtained from a magnetohydrodynamic (MHD) simulation.
However, a complete modeling of the evolution of the adiabatic particle distribution in a flare region has not been studied yet.
% So far, no one has shown the exact treatment of the evolution of the adiabatic particle distribution in flares, which should be understood through a direct comparison with observations. 

The correlation between the intensity of HXRs and the strength of convection electric fields has been reported by e.g., \cite{2002ApJ...565.1335Q} and \cite{2003ApJ...586..624A,2004ApJ...611..557A}.
Recently, \cite{2008ApJ...672L..69L} have observed the correlation between the hardness of the HXR spectrum and the strength of convection electric fields along flare ribbons. These observations indicate that the electric field greatly influences the distribution of electrons during their travel in the corona. The adiabatic model with actual coronal parameters can be tested through a direct comparison with these observations, because the model simultaneously describes particle acceleration, transport, and dissipation processes in a very wide area of the corona where particles are strongly magnetized.

To understand the particle acceleration, transport, and dissipation processes in solar flares, we theoretically study the evolution of the particle distribution based on the numerical simulation. For this purpose, we newly develop a comprehensive model based on the drift-kinetic theory, in which the evolution of the particle distribution function is described by means of a numerical simulation of the drift-kinetic Vlasov equation. The simulation can be performed with actual coronal parameters, allowing us to directly compare it with observations. In {\S}~\ref{sec:models} we describe equations of particles and electromagnetic fields in our model.
Simulation results are shown in {\S}~\ref{sec:calculation-results}.
The simulation identifies two dominant mechanisms of electron acceleration; the betatron acceleration at the top of closed loops and the inertia drift acceleration in open magnetic field lines. We discuss the results in {\S} \ref{sec:discussion}, and then conclude the paper in {\S} \ref{sec:conclusion-1}.

\section{Models}\label{sec:models}
The evolution of a collisionless plasma is fully described with the Vlasov equation in six-dimensional phase space (three dimensions in the configuration and velocity spaces). The equation describes the full kinetics of the particle transport along and across magnetic field lines, acceleration by electric fields, and the gyromotion, thus can resolve the inertia and gyro scales of particles. 
When the typical scale of interest is much larger than the gyro scale, it is rather convenient for focusing on the macroscopic distribution of particles to adopt a guiding-center equation (drift kinetics) than the full kinetic equation.
% In a solar flare, this is a reasonable approach because the gyro scale ($\sim 10^{-9} \; {\rm sec} \; {\rm and} \sim 10^{0} \; {\rm cm}$ for electrons) is much smaller than the flare scale ($\sim 10^{2} \; {\rm sec} \; {\rm and} \sim 10^{9} \; {\rm cm}$).
In this paper we treat the drift-kinetic Vlasov equation to study the time evolution of the particle distribution function following a macroscopic change of ambient fields.
\subsection{Basic Equations}\label{sec:basic-equations}
We begin with the relativistic guiding-center equations of motion \citep{1963RvGSP...1..283N},
\begin{eqnarray}
\dif{\vect r}{t} &=& {\vect v} = \frac{u_{\parallel}}{\gamma} {\vect b} + {\vect v}_{d}, \label{eq1}\\ 
\dif{u_{\parallel}}{t} &=& \gamma {\vect v}_{E} \cdot \dif{\vect b}{t} - \frac{M}{\gamma} \pdif{B}{s} + \frac{e}{m_0} E_{\parallel}, \label{eq2} \\
\dif{M}{t} &=& 0,\label{eq:1}\\
% {\vect v}_{d} &=& {\vect v}_{E} + 
% \frac{MB}{\gamma \omega_0} \frac{{\vect B} \times \nabla B}{B^2} + 
% \frac{1}{\omega_0}{\vect b} \times \left(u_{\parallel} \dif{\vect b}{t} + \gamma \dif{{\vect v}_{E}}{t}\right) \label{eq3},
{\vect v}_{d} &=& {\vect v}_{E} + 
\frac{MB}{\gamma \omega_0} {\vect b} \times \nabla \ln B + 
\frac{1}{\omega_0}{\vect b} \times \left(u_{\parallel} \dif{\vect b}{t} + \gamma \dif{{\vect v}_{E}}{t}\right) \label{eq3},
\end{eqnarray}
where $({\vect r}, {\vect v})$ is the guiding-center position and velocity of an individual particle, ${\vect v}_{d}$ is the drift velocity perpendicular to the magnetic field line, $e$ and $m_0$ are the charge and rest mass of a particle,
% \begin{eqnarray}
% && {\vect b} = \frac{\vect B}{B}, \;\;\; {\vect v}_{E} = \frac{{\vect E} \times {\vect B}}{B^2} c, \;\;\; E_{\parallel} = {\vect E} \cdot {\vect b}, \nonumber \\
% && {\vect u} = \gamma {\vect v}, \;\;\; u_{\parallel} = {\vect u} \cdot {\vect b}, \;\;\; \gamma = \left\{1-(v/c)^2\right\}^{-1/2} = \left\{(u/c)^2 + 1\right\}^{1/2}, \nonumber \\
% && \omega_0 = \frac{eB}{m_0 c}, \;\;\; M=\frac{u^2-u_{\parallel}^2}{2 B}, \nonumber \\
% && \dif{}{t} = \pdif{}{t} + \frac{u_{\parallel}}{\gamma}\pdif{}{s}+\left({\vect v}_{E} \cdot \nabla\right), \;\;\;  \pdif{}{s} = {\vect b} \cdot \nabla \nonumber,
% \end{eqnarray}
\begin{eqnarray}
&& {\vect b} = \frac{\vect B}{B}, \;\;\; {\vect v}_{E} = c \frac{{\vect E}}{B} \times {\vect b}, \;\;\; E_{\parallel} = {\vect E} \cdot {\vect b}, \nonumber \\
&& {\vect u} = \gamma {\vect v}, \;\;\; u_{\parallel} = {\vect u} \cdot {\vect b}, \;\;\; {\vect u}_{\perp}  = {\vect u} - u_{\parallel}{\vect b}, \nonumber \\
% && \gamma = \left\{(u/c)^2 + 1\right\}^{1/2}, \;\;\; \omega_0 = \frac{eB}{m_0 c}, \;\;\; M=\frac{u_{\perp}^2}{2 B}, \nonumber \\
&& \gamma = \mart{\left(u/c\right)^2+1}, \;\;\; \omega_0 = \frac{eB}{m_0 c}, \;\;\; M=\frac{u_{\perp}^2}{2 B}, \nonumber \\
&& \dif{}{t} = \pdif{}{t} + \frac{u_{\parallel}}{\gamma}\pdif{}{s}+\left({\vect v}_{E} \cdot \nabla\right), \;\;\;  \pdif{}{s} = {\vect b} \cdot \nabla \nonumber,
\end{eqnarray}
and $v_{d}$ is assumed to be much slower than the speed of light $c$.
In the drift-kinetic theory, dependent variables are reduced to $({\vect r},u_{\parallel},M)$ by the assumption of gyrotropy. In addition, $M$ is no longer a variable because of the conservation of the magnetic moment (eq. (\ref{eq:1})).

Since observable quantities are the energy and pitch-angle distribution of accelerated particles in the case of solar flares, we convert the variables $(u_{\parallel},M)$ into $(\gamma,\mu)$, where $\mu = \left({\vect v} \cdot {\vect b}\right)/v$ is the pitch-angle cosine of a particle.
%  measured in the fluid frame moving with ${\vect v}_{d}$.
 Their time derivatives are written as
\begin{eqnarray}
\dif{\gamma}{t} &=& \omega_0 \frac{{\vect v}_{d} \cdot {\vect E}}{c B} + \frac{MB}{\gamma c^2} \pdif{\ln B}{t},\label{eq4}\\
\dif{\mu}{t} &=& \dif{}{t}\left(\frac{u_{\parallel}}{u}\right) = \frac{1}{c \mart{\gamma^2-1}} \dif{u_{\parallel}}{t} - \frac{\mu \gamma}{\gamma^2 - 1} \dif{\gamma}{t},\label{eq5}
\end{eqnarray}
where we assume $E_{\parallel} = 0$. Using equations (\ref{eq2}) and (\ref{eq3}), equations (\ref{eq4}) and (\ref{eq5}) are rewritten as
% \begin{eqnarray}
% \dif{\gamma}{t} &=& \frac{\gamma^2-1}{\gamma} \frac{1-\mu^2}{2} \pdif{\ln B}{t} \nonumber \\
% &+& \frac{{\vect v}_{E}}{c} \cdot \left[ c \frac{\gamma^2-1}{\gamma} \left( \frac{1-\mu^2}{2} \nabla \ln B + \mu^2 \pdif{\vect b}{s}\right)
%  + \mu \mart{\gamma^2-1} \left(\pdif{}{t} + {\vect v}_{E} \cdot \nabla \right){\vect b} \right], \label{eq6}
% \end{eqnarray}
\begin{eqnarray}
\dif{\gamma}{t} = &\mart{\gamma^2-1}& \left[ \frac{\left(1-\mu^2\right) \mart{\gamma^2-1}}{2 \gamma} \left(\pdif{}{t} + {\vect v}_{E} \cdot \nabla \right) \ln B \right. \nonumber \\
&& \left. + \mu \frac{{\vect v}_{E}}{c} \cdot \left(\pdif{}{t} + \frac{\mu c \mart{\gamma^2-1}}{\gamma} \pdif{}{s} + {\vect v}_{E} \cdot \nabla \right) {\vect b} \right],\label{eq6}
\end{eqnarray}
% \begin{eqnarray}
% \dif{\mu}{t} = &(1-\mu^2)& \left[ -\frac{\mu}{2} \pdif{\ln B}{t} \right. \nonumber \\
% &&  + \frac{{\vect v}_{E}}{c} \cdot \left\{c \mu \left(-\frac{1}{2} \nabla \ln B + \pdif{\vect b}{s}\right) + \frac{\gamma}{\mart{\gamma^2 - 1}}\left(\pdif{}{t} + {\vect v}_{E} \cdot \nabla\right){\vect b}\right\} \nonumber \\
% && \left. - \frac{1}{2}\frac{c \mart{\gamma^2-1}}{\gamma} \pdif{\ln B}{s} \right].\label{eq7}
% \end{eqnarray}
\begin{eqnarray}
\dif{\mu}{t} = &\left(1-\mu^2\right)& \left[ -\frac{\mu}{2} \left(\pdif{}{t} + {\vect v}_{E} \cdot \nabla \right) \ln B \right. \nonumber \\
&& \left. + \frac{\gamma}{\mart{\gamma^2-1}}\frac{{\vect v}_{E}}{c} \cdot \left(\pdif{}{t} + \frac{\mu c \mart{\gamma^2-1}}{\gamma} \pdif{}{s} + {\vect v}_{E} \cdot \nabla \right) {\vect b} \right. \nonumber \\
&& \left. - \frac{1}{2}\frac{c \mart{\gamma^2-1}}{\gamma} \pdif{\ln B}{s} \right].\label{eq7}
\end{eqnarray}
The first term of the equations represents the betatron ($\nabla B$ drift) force, and the second term is the inertia drift { (general form of the curvature drift in the presence of electric fields)} force, respectively. The third term of equation (\ref{eq7}) is the magnetic mirror force. 
The polarization drift ($d{\vect v}_E / dt$ in eq. (\ref{eq3})) is negligible by the assumption $v_E/c \ll 1$.
% We ignore the polarization drift term ($d{\vect v}_E / dt$ in eq. (\ref{eq3})) by the assumption $v_E/c \ll 1$.

The time derivative of the position (eq. (\ref{eq1})) is expressed by the parallel and perpendicular drift motions to the magnetic field line. The drift motion consists of the $\exb$ drift ${\vect v}_{E}$, $\nabla B$ drift, inertia drift, and polarization drift (eq. (\ref{eq3})).
We estimate these drift velocities in a typical coronal condition. 
The $\exb$ drift is a motion of fluid described by MHD, up to an {\Alfven} velocity $\sim 2000 \; {\rm km \; s^{-1}}$. 
The $\nabla B$ and inertia drifts are $\sim \left(r_{g}/L\right)v$, where $r_{g} \sim 10^{0} \; {\rm cm}$ is the gyroradius and $L \sim 10^{9} \; {\rm cm}$ is the characteristic scale of magnetic fields. These are much slower than the {\Alfven} velocity. The polarization drift is further slower than these drifts.
 Therefore equation (\ref{eq1}) can be well approximated into
\begin{eqnarray}
\dif{\vect r}{t} &=& \frac{\mu c \mart{\gamma^2-1}}{\gamma} {\vect b} + {\vect v}_{E}.\label{eq:2}
\end{eqnarray}
Equations (\ref{eq6}) - (\ref{eq:2}) describe the motion of a particle with the guiding-center variables $({\vect r},\gamma,\mu)$. % Using these, we express the drift-kinetic Vlasov equation in the conservation form \citep[e.g.,][]{1984PhyS...29..141W}, for describing the time evolution of the distribution function of adiabatic particles,
% \begin{eqnarray}
% \pdif{f}{t}+\nabla \cdot \left(\dif{\vect r}{t}f\right) + \pdif{}{\mu}\left(\dif{\mu}{t} f\right) + \pdif{}{\gamma}\left(\dif{\gamma}{t} f\right) = 0,\label{eq:3}
% \end{eqnarray}
% where $f({\vect r},\gamma,\mu,t)$ is the distribution function defined as $f = dN/ \left(d{\vect r} d\gamma d\mu \right)$, $dN$ is the number of particles with $(\gamma,\mu)$, and $d{\vect r} d\gamma d\mu$ is the volume element in the phase space.

{ We define the particle distribution function $f({\bf r},\gamma,\mu) = dN/ \left(d{\vect r} d\gamma d\mu \right)$, where $dN$ is the number of particles with $(\gamma,\mu)$, and $d{\vect r} d\gamma d\mu$ is the volume element in the phase space. Its time evolution is described by the following continuity equation of the phase space density \citep[the drift-kinetic Vlasov equation; e.g.,][]{1984PhyS...29..141W,2005ipp..book.....G},
\begin{eqnarray}
\pdif{f}{t}+\nabla \cdot \left(\dif{\vect r}{t}f\right) + \pdif{}{\mu}\left(\dif{\mu}{t} f\right) + \pdif{}{\gamma}\left(\dif{\gamma}{t} f\right) = 0.\label{eq:3}
\end{eqnarray}
.}

\subsection{Electromagnetic Fields}\label{sec:electr-fields}
Equation (\ref{eq:3}) traces the orbit of the particle distribution function in ambient fields. For the determination of the fields, we employ the analytic model of two-dimensional magnetic fields of the flare proposed by \cite{1995SoPh..159..275L}. Their model is a superposition of potential and horizontal fields to impose an X-type neutral line at $(x,z) = (0,a)$, 
\begin{eqnarray}
A_y(x,z) &=& \left[\frac{z+d}{x^2+\left(z+d\right)^2} + \frac{z}{\left(a+d\right)^2}\right],\label{eq:4}\\
B_x(x,z) &=& -\pdif{A_y}{z} = -\left[\frac{x^2-\left(z+d\right)^2}{\left\{x^2+\left(z+d\right)^2\right\}^2} + \frac{1}{\left(a+d\right)^2}\right],\label{eq:5}\\
B_z(x,z) &=& \pdif{A_y}{x} = -\frac{2x\left(z+d\right)}{\left\{x^2+\left(z+d\right)^2\right\}^2},\label{eq:8}
\end{eqnarray}
% where%  $A_{y}$ is the flux function ($y$-component of the vector potential), 
where $d$ is the depth of a dipole moment. The $x$- and $z$-axes correspond to the tangential and normal directions relative to the solar surface, and the $y$-axis is along the neutral line, respectively. Note that the above formulae are slightly different from the original ones, because we take the origin at the surface whereas \citeauthor{1995SoPh..159..275L} took it at the neutral line.

\begin{figure}[htbp]
\centering
\epsscale{0.5}
\plotone{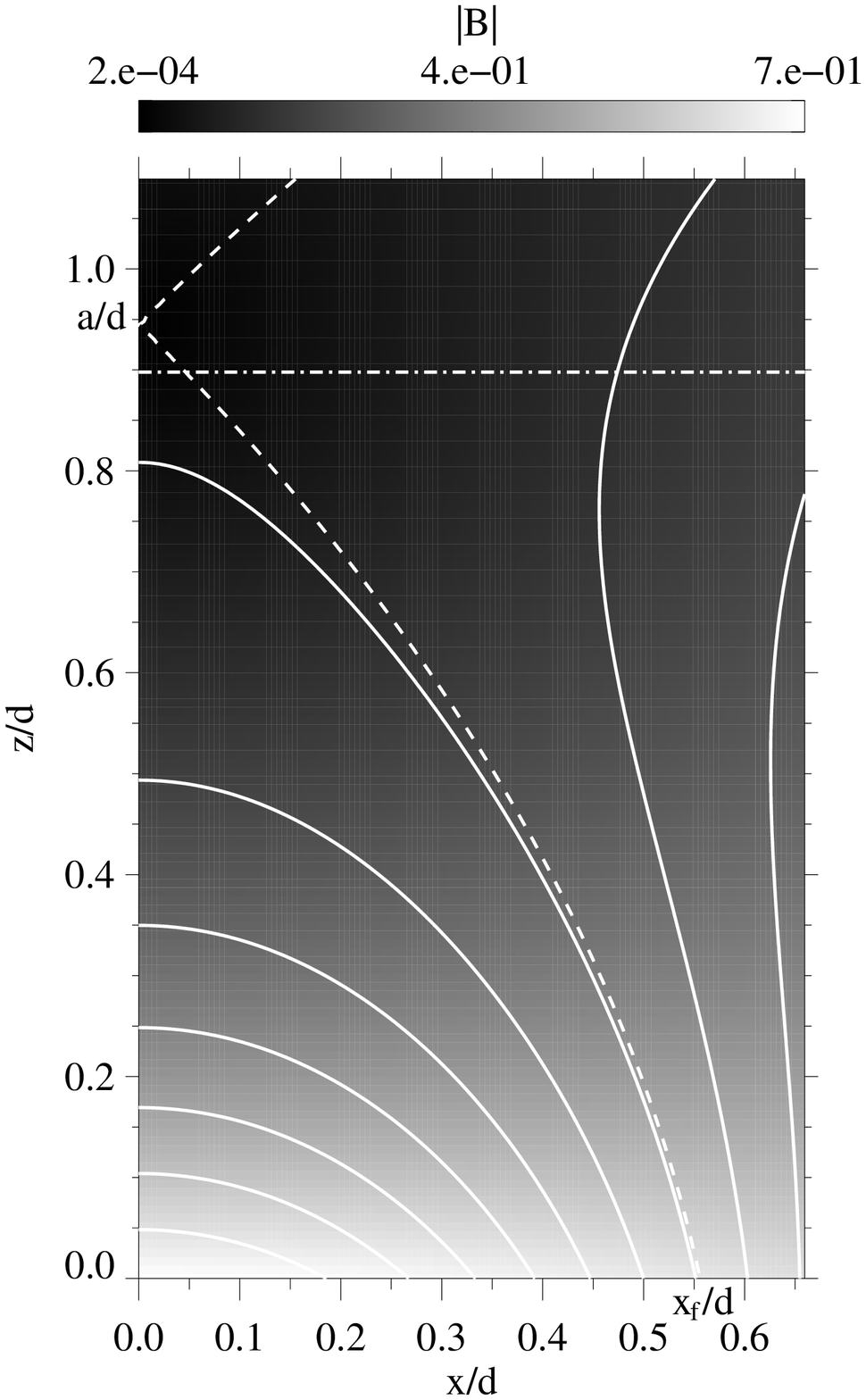}
\caption{Magnetic field configuration (eq. (\ref{eq:4})) with an aspect ratio $R=a/x_f=1.7$. The white solid lines are the field lines, and the color contour is the field strength. The dashed line is the magnetic separatrix. The dash-dotted line is the upper boundary of the simulation domain.}
\label{fig:linmodel}
\end{figure}

We introduce an $x$-position of the footpoint of the magnetic separatrix $x_{f}$, which is magnetically connected to the neutral line. The relation $A_y(0,a)=A_y(x_f,0)$ yields
% \begin{eqnarray}
% \frac{a}{d}&=&\left(\frac{x_f}{d}+\mart{\left(\frac{x_f}{d}\right)^2+1}\right)\frac{x_f}{d}, \label{eq:6}\\
% \frac{x_f}{d}&=&\mart{\frac{1}{\left(d/a\right)^2+2\left(d/a\right)}}.\label{eq:7}
% \end{eqnarray}
\begin{eqnarray}
\frac{a}{d}&=&\left(\frac{x_f}{d}+\mart{\left(\frac{x_f}{d}\right)^2+1}\right)\frac{x_f}{d}. \label{eq:6}
\end{eqnarray}
We further introduce an aspect ratio of the magnetic field configuration $R = a/x_f$, which yields
% \begin{eqnarray}
% R = \frac{a}{x_f} = \frac{x_f}{d} + \mart{\left(\frac{x_f}{d}\right)^2+1},\label{eq:9}
% \end{eqnarray}
% yielding
\begin{eqnarray}
\frac{x_f}{d} = \frac{R^2-1}{2R}.\label{eq:10}
\end{eqnarray}
The geometry of the magnetic field is expressed as a function of $R$, which is treated as one of our simulation parameters. Observationally, \cite{1996ApJ...470.1198A,1996ApJ...468..398A} have statistically estimated the ratio between the height of the acceleration site and the half length of the footpoint distance from a time-of-flight analysis to HXR data, giving $1.7 \pm 0.4$. The magnetic field configuration with $R=1.7$ is shown in Figure \ref{fig:linmodel}.

The flare evolution can be characterized by the temporal change of the magnetic field configuration. Flares frequently show a separating motion of ribbons in wavelengths sensitive to the chromosphere such as H$\alpha$ \citep[e.g.,][]{2002ApJ...565.1335Q,2003ApJ...586..624A,2004ApJ...611..557A,2009ApJ...697..843M}.
 Based on the magnetic reconnection model, it is interpreted as a chromospheric counterpart of the rising motion of the neutral line. \cite{1992PASJ...44L..63T} reported a growth of a cusp-shaped loop, increasing its height with time. To include the evolution in our simulation, we change the footpoint position of the magnetic separatrix with time. The time derivative (apparent velocity) of the footpoint position is given as a Gaussian profile,
\begin{eqnarray}
\dif{x_{f}}{t} = v_{p} \exp\left[-\frac{1}{2} \left(\frac{t-t_{p}}{\tau}\right)^{2}\right].\label{eq:11}
\end{eqnarray}
% where $v_{p}$ is the footpoint velocity at the flare peak time $t_{p}$, and $\tau$ is the duration. 
From equations (\ref{eq:6}) and (\ref{eq:10}), $a$ and $R$ also change with time.

The temporal change of magnetic fields induces electric fields. Since the electrostatic potential can be assumed to be zero in the two-dimensional system, the $y$-component of the electric field is determined from the Faraday's law,
\begin{eqnarray}
E_y &=& -\frac{1}{c} \frac{\partial A_{y}}{\partial t} = -\frac{1}{c} \frac{d x_{f}}{dt}\pdif{a}{x_f}\pdif{A_y}{a} \nonumber \\
&=& \left[\frac{2x_f}{d}+\frac{2 \left(x_f/d\right)^2+1}{\mart{\left(x_f/d\right)^2+1}}\right] \frac{2z}{\left(a+d\right)^3} \frac{1}{c} \frac{d x_{f}}{dt},\label{eq:12}
\end{eqnarray}
and other components are zero.
The model electric field is a linear function of $z$, taking zero at the surface. This means that the magnetic field lines at the surface are stationary.
%  do not change their position $(v_{E}(z=0) = 0)$.
%  Since it is proportional to $dx_f/dt$, the electric field peaks at $t = t_p$.
Since the electric field is proportional to $dx_f/dt$, the strength grows till $t = t_p$ (we call the period as the rising phase), and then is reduced (the declining phase).

% Note that in equation (\ref{eq:11}) we move the position of the magnetic separatrix, not magnetic field lines at the solar surface. 
\subsection{Simulation Setup}\label{sec:init-bound-cond}
Using the listed equations, we numerically solve the drift-kinetic Vlasov equation (\ref{eq:3}). Since we assume two-dimensionality in the configuration space, the equation is reduced to a four-dimensional compressive advection equation. To solve it, we adopt the R-CIP-CSL2 scheme with operator splitting technique \citep{2001JCoPh.174..171N}, which simultaneously solves the integrated values of $f$ as well as $f$ itself to keep ``subgrid'' information and to satisfy mass conservation. 

Simulation parameters are $d=1.5 \times 10^{9} \; {\rm cm}$ and $ R(t=0) = 1.7$.
% Considering that the position of flare ribbons corresponds to the root of the magnetic separatrix, we set the parameters in equation (\ref{eq:11}) based on observations \citep[e.g.,][]{2002ApJ...565.1335Q};
%  $v_{p} = 60 \; {\rm km \; s^{-1}}$, $\tau = 1.77 \; {\rm s}$, and $t_{p} = 5.0 \; {\rm s}$.
Considering that the position of flare ribbons corresponds to the root of the magnetic separatrix, we set the parameters in equation (\ref{eq:11}) based on observations. { \cite{2002ApJ...565.1335Q} have observed the apparent motion of flare kernels seen in the H$\alpha$ line. The velocity reaches several ten to 100 ${\rm km \; s^{-1}}$ within a time scale of $\sim 10$ s (see Figs. 5 and 6 in their paper). Referring to the first spike in Figure 5(b) of \citeauthor{2002ApJ...565.1335Q}, we set}
 $v_{p} = 60 \; {\rm km \; s^{-1}}$, $\tau = 1.77 \; {\rm s}$, and $t_{p} = 5.0 \; {\rm s}$.
The simulation runs till $t=10$ s. Length $(x,z,a,x_{f})$ is normalized to $d$. % The simulation is performed with actual parameters of the solar corona.

\begin{figure}[htbp]
\centering
\epsscale{0.6}
\plotone{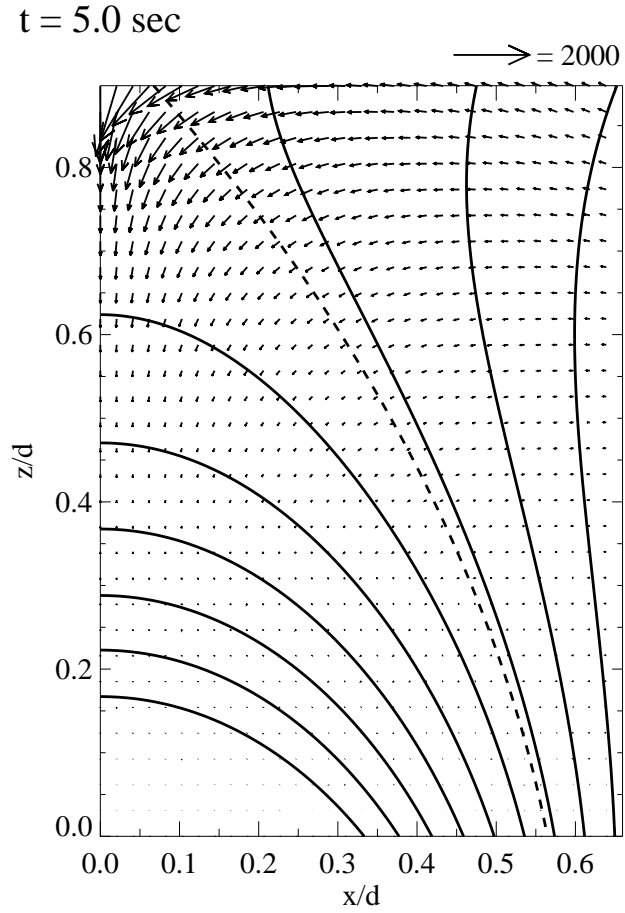}
\caption{$\exb$ drift velocity field distribution calculated with eqs. (\ref{eq:5}), (\ref{eq:8}), and (\ref{eq:12}) at $t = 5$ s. The unit of arrows is in ${\rm km \; s^{-1}}$. The black lines are the magnetic field lines, and the dashed line is the magnetic separatrix.}
\label{fig:linmodel_vel}
\end{figure}

We limit a calculation domain of the configuration space to the area below the dash-dotted line in Figure \ref{fig:linmodel}, in which the $\exb$ drift velocity calculated with equations (\ref{eq:5}), (\ref{eq:8}), and (\ref{eq:12}) does not exceed a typical {\Alfven} velocity in the corona $(\sim 2000 \; {\rm km \; s^{-1}})$. In other word, our model can not properly describe the electric field around the neutral line, where the $\exb$ drift velocity is calculated to be an unrealistic value.
Figure \ref{fig:linmodel_vel} shows an example of the $\exb$ drift velocity field distribution within the simulation domain.
%  The velocity shows a lower-left direction.

Initially, 3 keV isotropic electrons are uniformly distributed in the configuration space. 
{ We will discuss the assumption of the relatively high initial temperature in \S~\ref{sec:discussion}.}
The initial uniform distribution is reasonable, because the hydrostatic scale height is much longer than the size of the simulation domain. 
At $x=x_{\rm max}$ and $z=z_{\rm max}$ we impose the open boundary condition where incoming fluxes are assumed while outgoing fluxes are perfectly lost. The incoming flux is fixed to the initial distribution. We also impose another open boundary condition at $z=0$, where a cold dense plasma is fixed to exclude the possibility of hot electrons originating below the surface.
%  The boundary condition at $x=0$ is symmetric; $\partial\left\{f(\mu)-f(-\mu)\right\}/\partial x=0$. 
 The boundary condition at $x=0$ is symmetric: $f(x,z,\gamma,\mu) = f(-x,z,\gamma,-\mu)$.
We note that the initial uniform and isotropic distribution is not a steady state solution in our model, because of the different boundary conditions at $z=z_{\rm max}$ (hot tenuous plasma) and $z=0$ (cold dense plasma).

We use $64\times96$ cells in $(x,z)$ space, and $64\times64$ cells in $(\gamma,\mu)$ space. A domain of $\gamma$ is 1.01-1.20, corresponding to 0.511-102.2 keV electrons. The grids in $\gamma$ space are logarithmically spaced.

\section{Simulation Results and Interpretation}\label{sec:calculation-results}
\begin{figure}[htbp]
\centering
\epsscale{0.9}
\plotone{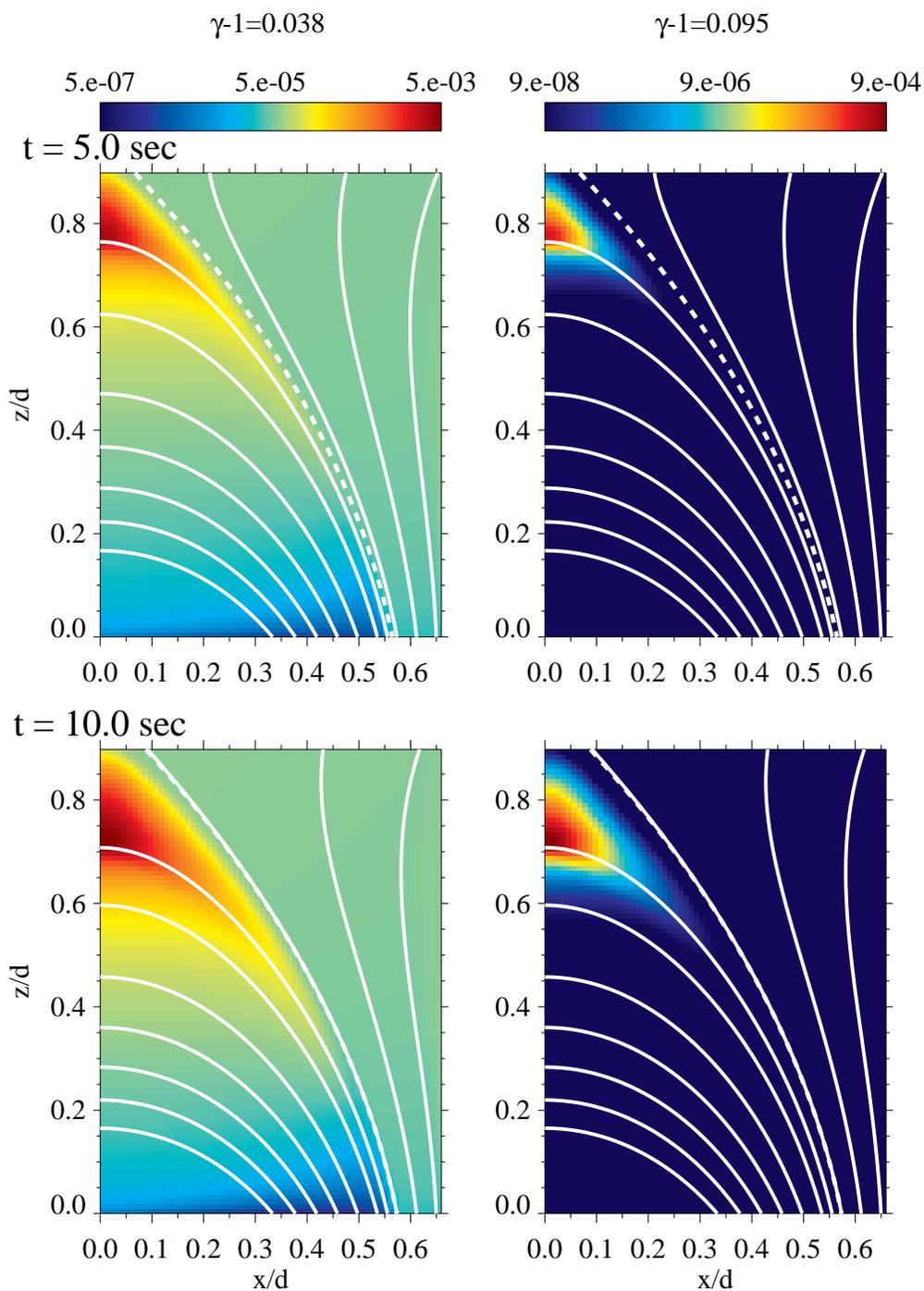}
\caption{Spatial distribution of the number of 20 keV (left) and 50 keV (right) omnidirectional electrons at $t = 5 \; {\rm s}$  (top) and $10 \; {\rm s}$ (bottom). The white lines are the magnetic field lines, and the dashed lines are the magnetic separatrix.}
\label{fig:xzimgs}
\end{figure}

Several snapshots of our simulation results are in Figure \ref{fig:xzimgs}, in which the spatial distributions of the number of 20 keV and 50 keV omnidirectional electrons at $t = 5$ s and 10 s are presented.
%  The color scale in both Figures is fourth order of magnitude smaller than the maximum value. 
In both energies the electron number increases significantly around the top of closed loops.
The altitude of the peak position of the electron number decreases with time, due to the shrinkage of the loops.
The distributions in the loops diffuse along the magnetic field lines. The spatial diffusion is more prominent for lower energy electrons.
Because of the boundary condition at $z=0$, the number of electrons decreases at the bottom area.

\subsection{Electron Distribution in Closed Loops}\label{sec:electr-distr-clos}
\begin{figure}[htbp]
\centering
\epsscale{1.0}
\plotone{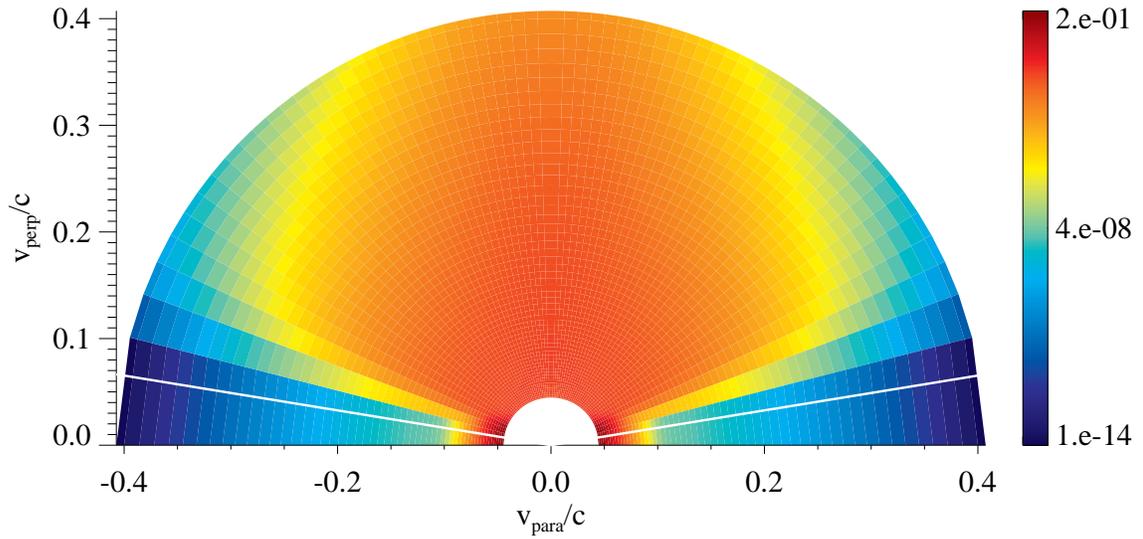}
\caption{Velocity space distribution at $(x,z)=(0,0.7)$ at $t=10 \; {\rm s}$, corresponding to the top of a closed loop. The horizontal and vertical axes correspond to the velocity parallel and perpendicular to the magnetic field line. The white lines denote the loss-cone angle.}
\label{fig:psd_lt}
\end{figure}

As seen in Figure \ref{fig:xzimgs}, a strong acceleration takes place around the loop top, at which the fastest $\exb$ drift velocity is observed (see Fig. \ref{fig:linmodel_vel}). The velocity space distribution is of help to understand the mechanism around there. Figure \ref{fig:psd_lt} shows the velocity space distribution function taken at the loop top $(x,z)=(0,0.7)$ at $t=10$ s. The horizontal and vertical axes correspond to the velocity parallel and perpendicular to the magnetic field line.

\begin{figure}[htbp]
\centering
\epsscale{0.9}
\plotone{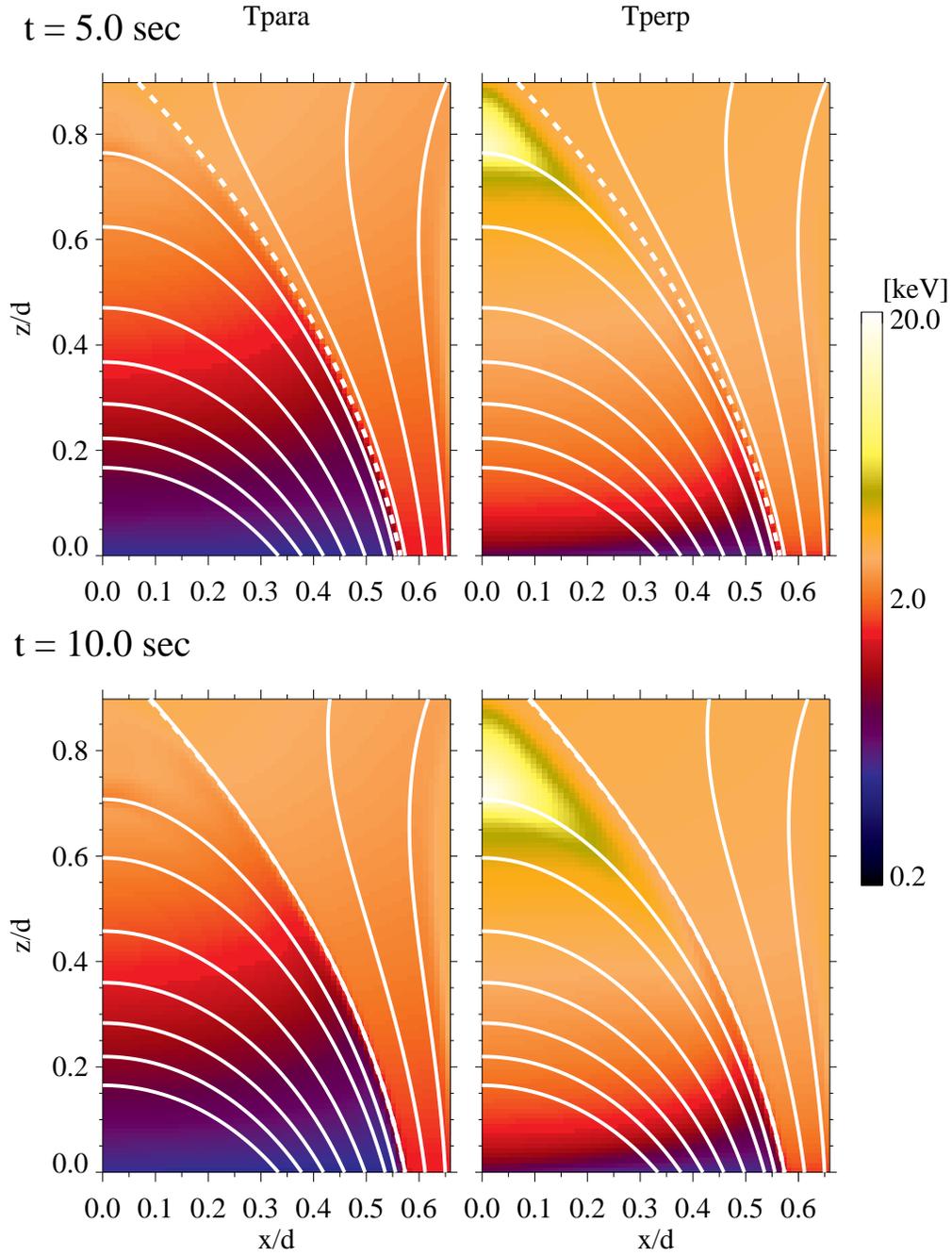}
\caption{Spatial distribution of the parallel (left) and perpendicular (right) temperatures at $t = 5$ s (top) and 10 s (bottom), calculated from eqs. (\ref{eq:16}) and (\ref{eq:14}). The white lines are the magnetic field lines, and the dashed lines are the magnetic separatrix.}
\label{fig:temp_dist}
\end{figure}

The distribution significantly deviates from the initial, isotropic Maxwell distribution.
It shows the loss-cone distribution with a loss-cone angle (white lines) defined as $\alpha = \sin^{-1} \mart{{B(x,z)}/{B(x_{FP},0)}}$,
% \begin{eqnarray}
% \alpha = \sin^{-1} \mart{\frac{B(x,z)}{B(x_{FP},0)}},\label{eq:13}
% \end{eqnarray} 
where $x_{FP}$ is the $x$-position at the footpoint of a specific field line determined from $A_{y}(x_{FP},0) = A_{y}(x,z).$
The distribution also shows a strong enhancement of the number of electrons perpendicular to the field line. 
The perpendicular enhancement is energy dependent; the distribution is more anisotropic for higher energy electrons. This means that high energy electrons are produced by the acceleration perpendicular to the field line, that is, the betatron ($\nabla B$ drift) acceleration.
% The betatron force appears in the first term of equations  (\ref{eq6}) and (\ref{eq7}).
Since the electron pitch angle becomes close to 90 degrees by the betatron acceleration, accelerated electrons are further confined and energized at the loop top with time.
By taking the second-order momentum of the distribution function, we estimate the average energies (temperatures) parallel and perpendicular to the magnetic field line as
\begin{eqnarray}
T_{\parallel}(x,z,t) &=& \inty{1}{\infty} d\gamma \inty{-1}{1} d\mu \frac{u_{\parallel}^2}{\gamma} f,\label{eq:16}\\
T_{\perp}(x,z,t) &=& \frac{1}{2} \inty{1}{\infty} d\gamma \inty{-1}{1} d\mu \frac{u_{\perp}^2}{\gamma} f,\label{eq:14}
\end{eqnarray}
where $u_{\parallel}^2 = \mu^2 \left(\gamma^2-1\right)$ and $u_{\perp}^2 = \left(1-\mu^2\right) \left(\gamma^2-1\right)$.
The spatial distributions of the parallel and perpendicular temperatures at $t = 5$ s and 10 s are shown in Figure \ref{fig:temp_dist}. Around the loop top, we can see a slight decrease of the parallel temperature and a strong increase of the perpendicular temperature relative to an initial temperature 3 keV.
The decrease of the parallel temperature is due to the loss of electrons inside the loss cone.
% The increase of the perpendicular temperature is resulted from the betatron ($\nabla B$ drift) acceleration.
% % The betatron force appears in the first term of equations  (\ref{eq6}) and (\ref{eq7}).
% Since the betatron acceleration increases the electron pitch angle (see the first term of eq. (\ref{eq7})), accelerated electrons are further confined and energized at the loop top with time.
% As a result, the distribution of higher energy electrons is more enhanced perpendicular to the field line.
The perpendicular temperature reaches a maximum $\sim 15$ keV at $(x,z) = (0,0.7)$ at $t = 10$ s. 
The acceleration increases the perpendicular temperature by a factor of 5 relative to the initial.
% From the first term of equation (\ref{eq6}) the rate of the betatron acceleration for electrons with $\mu = 0 \; (u_{\parallel} = 0)$ is equal to the compression ratio of the magnetic field strength;
% \begin{eqnarray}
% && \frac{1}{\epsilon}\dif{\epsilon}{t} \simeq \pdif{\ln B}{t} + \vect{v}_{E} \cdot \nabla \ln B = \frac{1}{B} \dif{B}{t},\nonumber \\
% &\Rightarrow& \frac{\epsilon_2}{\epsilon_1} = \frac{B_2}{B_1},
% \end{eqnarray}

On the other hand, the rate of the betatron acceleration for electrons with $\mu = 0$ is easily obtained from the conservation of the magnetic moment (eq. (\ref{eq:1})),
\begin{eqnarray}
\frac{\epsilon_2}{\epsilon_1} = \frac{B_2}{B_1},\label{eq:25}
\end{eqnarray}
where $\epsilon = \gamma-1 (\ll 1)$ is the kinetic energy.
 Since the loop at $(x,z)=(0,0.7)$ at $t = 10$ s is convected from the initial position $(x,z)=(0,0.89)$ by the $\exb$ drift, a compression ratio at the loop top is
\begin{eqnarray}
\frac{B(t=10 \; {\rm s})}{B(t=0 \; {\rm s})} \sim \frac{0.0935}{0.0156} = 6,\label{eq:15}
\end{eqnarray}
which is larger than the estimated increase rate of the perpendicular temperature. This is because the betatron acceleration less works for electrons with smaller pitch angles. They are accelerated rather parallel to the magnetic field line through the inertia drift. 

To consider the acceleration due to the inertia drift in the closed loop, it is convenient to introduce the longitudinal adiabatic invariant, $u_{\parallel} L = {\rm constant}$.
% The rate of the inertia drift acceleration in closed loops is obtained from the conservation of the longitudinal adiabatic invariant, $u_{\parallel} L$.
Here $L$ is the travel length of electrons measured along the field line,
\begin{eqnarray}
L &=& 2 \inty{\rm loop-top}{\rm mirror} ds = 2 \inty{0}{x_{\rm mirror}} \mart{1+\left(B_{z}/B_{x}\right)^2}dx, \label{eq:26}
\end{eqnarray}
where $x_{\rm mirror}$ is the $x$-position at the mirror point. For electrons with the pitch angle close enough to the loss-cone angle, $x_{\rm mirror} = x_{FP}$.
The inertia drift acceleration in the loop is expressed as the increase of the parallel energy due to the decrease of the length of the shrinking loop. The rate of the inertia drift acceleration is 
\begin{eqnarray}
\frac{\epsilon_2}{\epsilon_1} = \left(\frac{L_1}{L_2}\right)^2.\label{eq:27}
\end{eqnarray}
By numerically calculating the length of the loop that is initially at $(x,z) = (0,0.89)$ (the same loop as in eq. (\ref{eq:15})), we obtain
\begin{eqnarray}
\left(\frac{L(t=0 \; {\rm s})}{L(t=10 \; {\rm s})}\right)^2 \sim \left(\frac{1.076}{0.943}\right)^2 = 1.3.\label{eq:28}
\end{eqnarray}
Electrons that experience the inertia drift acceleration escape from the loop top and their distribution diffuses along the field line.
This yields the perpendicular temperature that is not as much as expected from the field compression (eq. (\ref{eq:15})). 
Nonetheless, the resulting velocity space distribution is enhanced perpendicular to the field line (Fig. \ref{fig:psd_lt}), because the rate of the perpendicular acceleration $\sim 6$ is larger than the parallel $\sim 1.3$.
{This is the general feature of acceleration with conserving both the magnetic moment and longitudinal invariant.} 
% Since the the efficiency of these accelerations depends on an ambient field configuration (eqs. (\ref{eq6}) and (\ref{eq7})), the resulting electron distribution is controlled by the field configuration.
% The spatial distribution as well as the velocity space distribution of electrons are determined by the efficiency of these accelerations, which depends on an ambient field configuration (eqs. (\ref{eq6}) and (\ref{eq7})). 

% In our flare model, the perpendicular acceleration is much efficient than the parallel acceleration in the closed loop. Majority of electrons are enhanced perpendicular to the field line and are confined at the loop top. This is also understood by comparing the spatial distribution of electrons with different energies (Figs. \ref{fig:xzimg_20kev} and \ref{fig:xzimg_49kev}), in which higher energy electrons are more confined at the loop top. This means that high energy electrons are produced mainly by the perpendicular betatron acceleration mechanism. 
 
\subsection{Electron Distribution in Open Field lines}\label{sec:electr-distr-open}
\begin{figure}[htbp]
\centering
\epsscale{1.0}
\plotone{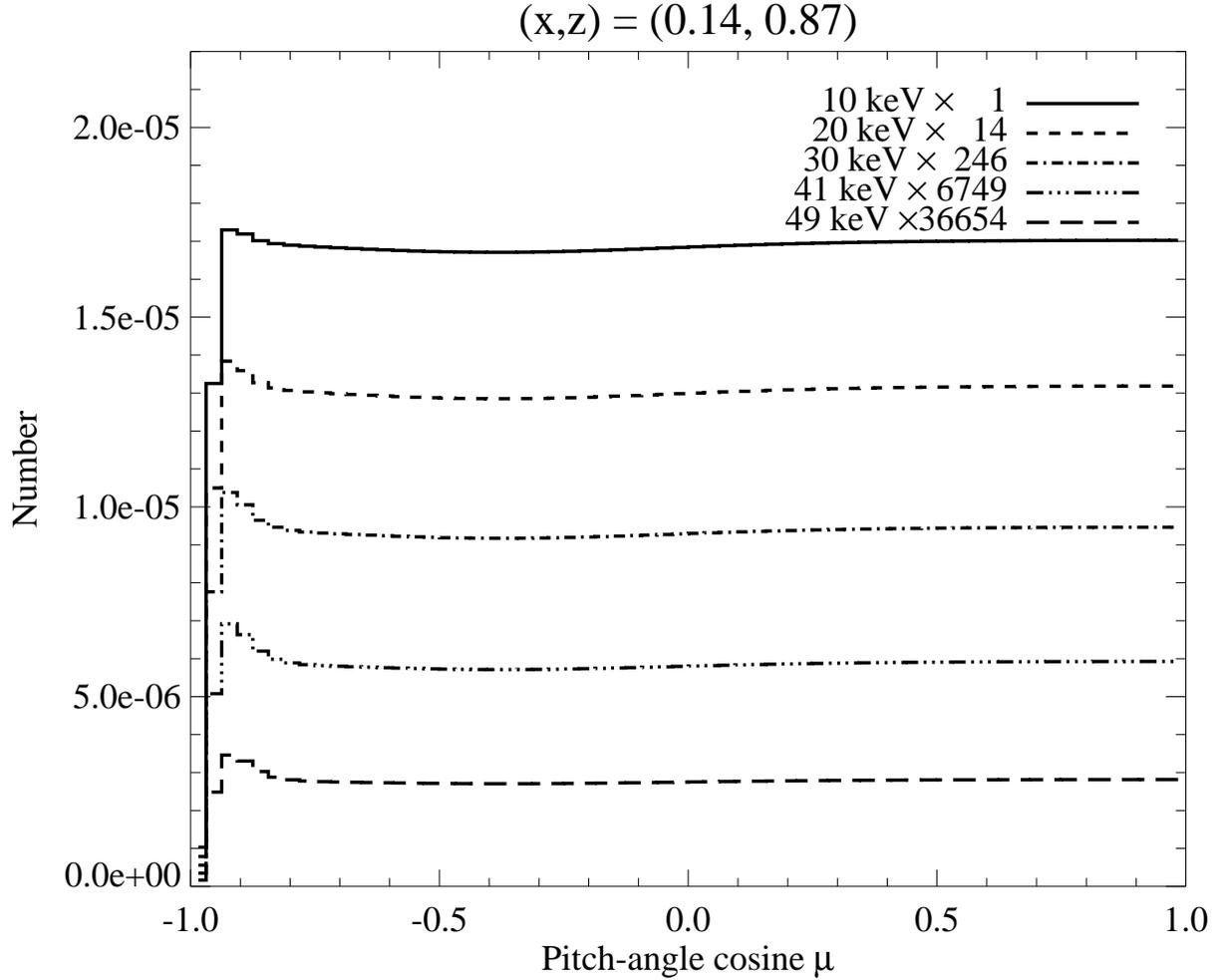}
\caption{Electron pitch-angle distribution in the open field line near the magnetic separatrix $(x,z) = (0.14,0.87)$ at $t = 5$ s. The positive (negative) $\mu$ corresponds to the sunward (antisunward) direction. Different lines represent the electrons with different energies. Number density is multiplied by factors for illustration, as shown in the legend on the upper-right corner.}
\label{fig:psd_sp}
\end{figure}

In the previous section, we discussed the acceleration taking place in the closed loop. It is interesting to discuss whether the acceleration occurs also in open field lines.
%  In Figures \ref{fig:xzimg_20kev} and \ref{fig:xzimg_49kev}, any mechanism does not seem to work in open field lines (right side region).
%  With a careful check of the velocity space distribution, however, we find a weak acceleration of upward electrons in open magnetic field lines.
Figure \ref{fig:psd_sp} shows the electron pitch-angle distribution in the open field line near the magnetic separatrix at $(x,z) = (0.14,0.87)$ at $t = 5$ s. The positive (negative) value of the pitch-angle cosine corresponds to the sunward (antisunward) direction. 
% The number density of electrons with different energies are drawn by multiplying factors, for illustration. 

The pitch-angle distribution is different between sunward and antisunward directions. For sunward ($\mu > 0$) the distribution is almost isotropic, because isotropic electrons are continuously injected from the upper boundary. For antisunward ($\mu < 0$), on the other hand, the distribution deviates from an isotropic distribution. We see lack of electrons with $\mu \sim -1$, also due to the boundary condition. Since we impose the lower boundary condition as almost vacuum for high energy electrons, the antisunward electrons are composed by those that are injected from the upper boundary (with $\mu > 0$) and then are reflected (change the sign of $\mu$) by the magnetic mirror force. 
Injected electrons with $\mu \sim 1$ can not be reflected, and are lost at the lower boundary. Therefore electrons with $\mu \sim -1$ are not found along open field lines.

We find a slight enhancement of the number of electrons with $\mu \sim -0.95$ and decrease with $-0.8 \lsim \mu < 0$, compared to those with $\mu > 0$. 
% We find a slight enhancement of the number of electrons with $\mu \sim -0.95$, compared to that with $\mu > 0$.
This is an evidence of the parallel acceleration to the magnetic field line due to the inertia drift: Electrons are accelerated antisunward along the curved line by the centrifugal force. % The inertia drift force appears in the second term of equations (\ref{eq6}) and (\ref{eq7}).
 This mechanism has been proposed for the generation of energetic particles in the planetary magnetosphere \cite[e.g.,][]{1996JGR...10117409D,2005AnGeo..23.3389D}, and in quasi-perpendicular shocks \citep{2007ApJ...661..190A}.

According to \cite{2007ApJ...661..190A}, we discuss the resulting pitch-angle distribution of this acceleration. In the rest frame, electrons gain energy from convection electric fields through the inertia drift. 
It is convenient to transform the frame into the so-called de Hoffmann-Teller frame moving with $\vect{v}_{E}$ relative to the rest frame, in which the electric field vanishes and hence electrons do not gain energy from convection electric fields.
In this frame, however, magnetic mirror points move relative to electrons when a magnetic field line has curvature. The velocity of the mirror point measured in the de Hoffmann-Teller frame is 
\begin{eqnarray}
\vect{v}_{\rm mirror} = - \vect{v}_{E} \tan \theta,\label{eq:17} 
\end{eqnarray}
where $\theta$ is an angle of the magnetic field line between the departure and mirror points of an electron. When $\vect{v}_{E}$ is sunward in the rest frame, the mirror point moves antisunward in the de Hoffmann-Teller frame. Sunward electrons experience a head-on reflection with the mirror points, which increases the electron parallel velocity by $2 v_{\rm mirror}$ and changes the direction.
%  This situation is realized in the model electromagnetic fields (Fig. \ref{fig:linmodel_vel}). 

The changes of the electron pitch angle and speed by the reflection are written as
\begin{eqnarray}
\mu' &=& - \left(\frac{\mu v + 2v_{\rm mirror}}{v'}\right), \;\;\; {\rm (if \; \mu > 0)},\label{eq:18}\\ 
(v')^2 &=& \left(\mu' v'\right)^2 + (1-\mu^2)v^2,\label{eq:21}
\end{eqnarray}
where variables with (without) a prime are measured after (before) the reflection. The second term of equation (\ref{eq:21}) is the square of the perpendicular velocity, which remains constant throughout the reflection. The resulting antisunward electron distribution $f'$ is 
\begin{eqnarray}
f'(1+\epsilon',\mu') = f_{\rm sunward}(1+\epsilon,\mu), \;\;\; {\rm (if \; \mu' < 0)}, \label{eq:23}
\end{eqnarray}
where $\epsilon = \gamma-1 \simeq (v/c)^2/2$.
% That the sunward electron distribution $f$ is an almost isotropic Maxwellian yields the second equality.
Note that the sunward electron distribution $f_{\rm sunward}$ is independent of $\mu$, $f_{\rm sunward}(1+\epsilon,\mu) = f_{\rm sunward}(1+\epsilon)$, because it is an almost isotropic Maxwellian.
With equations (\ref{eq:18}) and (\ref{eq:21}), equation (\ref{eq:23}) results in
\begin{eqnarray}
f'(1+\epsilon',\mu') = f_{\rm sunward}(1+\epsilon'+4\mu' \mart{\epsilon' \epsilon_{\rm mirror}}+4\epsilon_{\rm mirror}),\label{eq:19}
\end{eqnarray}
where $\epsilon_{\rm mirror} = (v_{\rm mirror}/c)^2/2$.
%  If $\epsilon_{\rm mirror}$ is known, we can calculate equation (\ref{eq:19}).

To determine $\epsilon_{\rm mirror}$, we suppose that an electron with $\mu_0 (> 0)$ is initially at $\vect{S}$, moves sunward, and then is reflected at $\vect{R}$. The position of the mirror point is determined from the conservation of the magnetic moment,
\begin{eqnarray}
B(\vect{R}) = \frac{B(\vect{S})}{1-\cos \mu_0^2},\label{eq:20}
\end{eqnarray}
which depends on the electron initial condition.
The velocity of the mirror point (eq. (\ref{eq:17})) depends on the electron initial condition as well as the field configuration, which is evaluated as
\begin{eqnarray}
\vect{v}_{\rm mirror} = - \vect{v}_{E}(\vect{S}) \tan \left[ \cos^{-1} \left\{ \vect{b}(\vect{S}) \cdot \vect{b}(\vect{R})\right\}\right].\label{eq:22}
\end{eqnarray}
We need to determine the mirror point of electrons with different pitch angles, to calculate equation (\ref{eq:19}) with (\ref{eq:22}).

\begin{figure}[htbp]
\centering
\epsscale{1.0}
\plotone{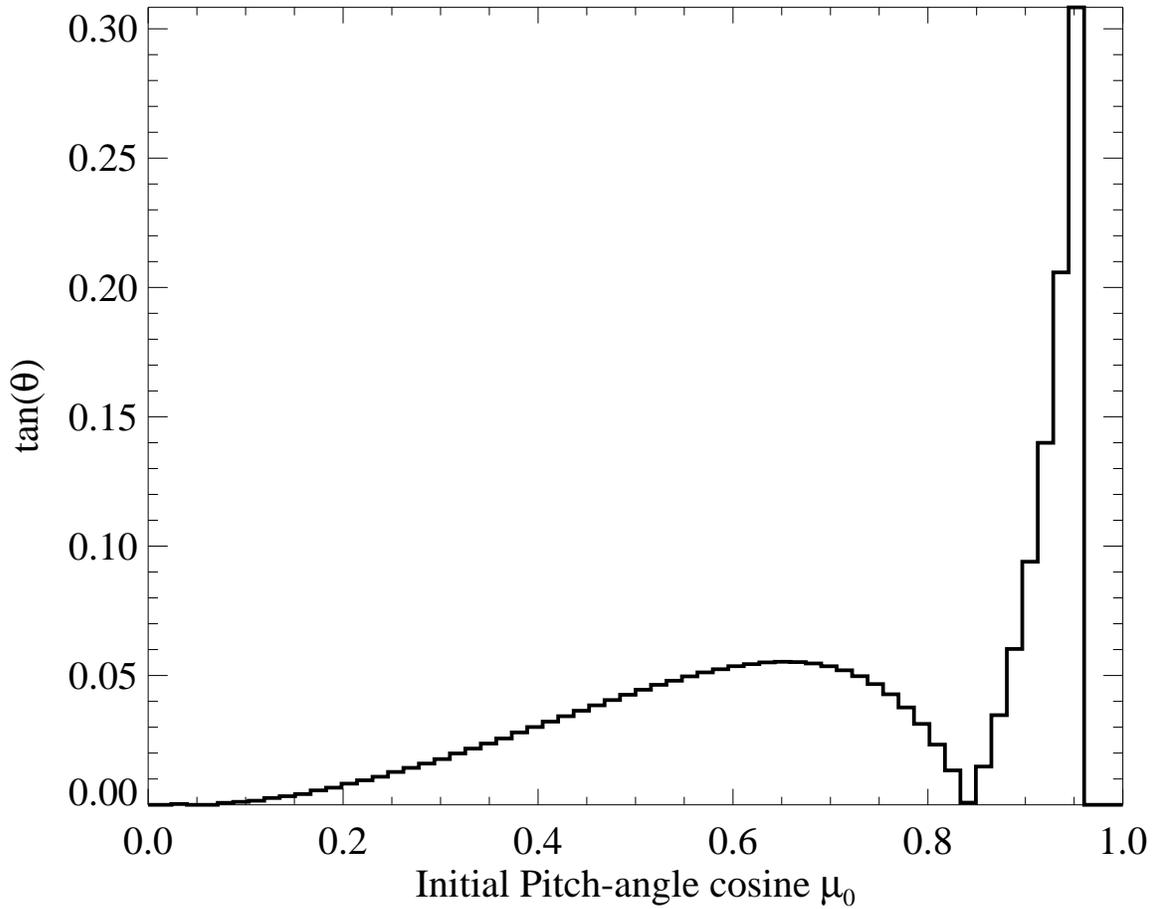}
\caption{Tangent of an angle of the magnetic field line between the departure and mirror points of electrons, as a function of their initial pitch-angle cosine (see eqs. (\ref{eq:20}) and (\ref{eq:22})), obtained from a test particle simulation with eqs. (\ref{eq6}) - (\ref{eq:2}).}
\label{fig:tant}
\end{figure}

\begin{figure}[htbp]
\centering
\epsscale{1.0}
\plotone{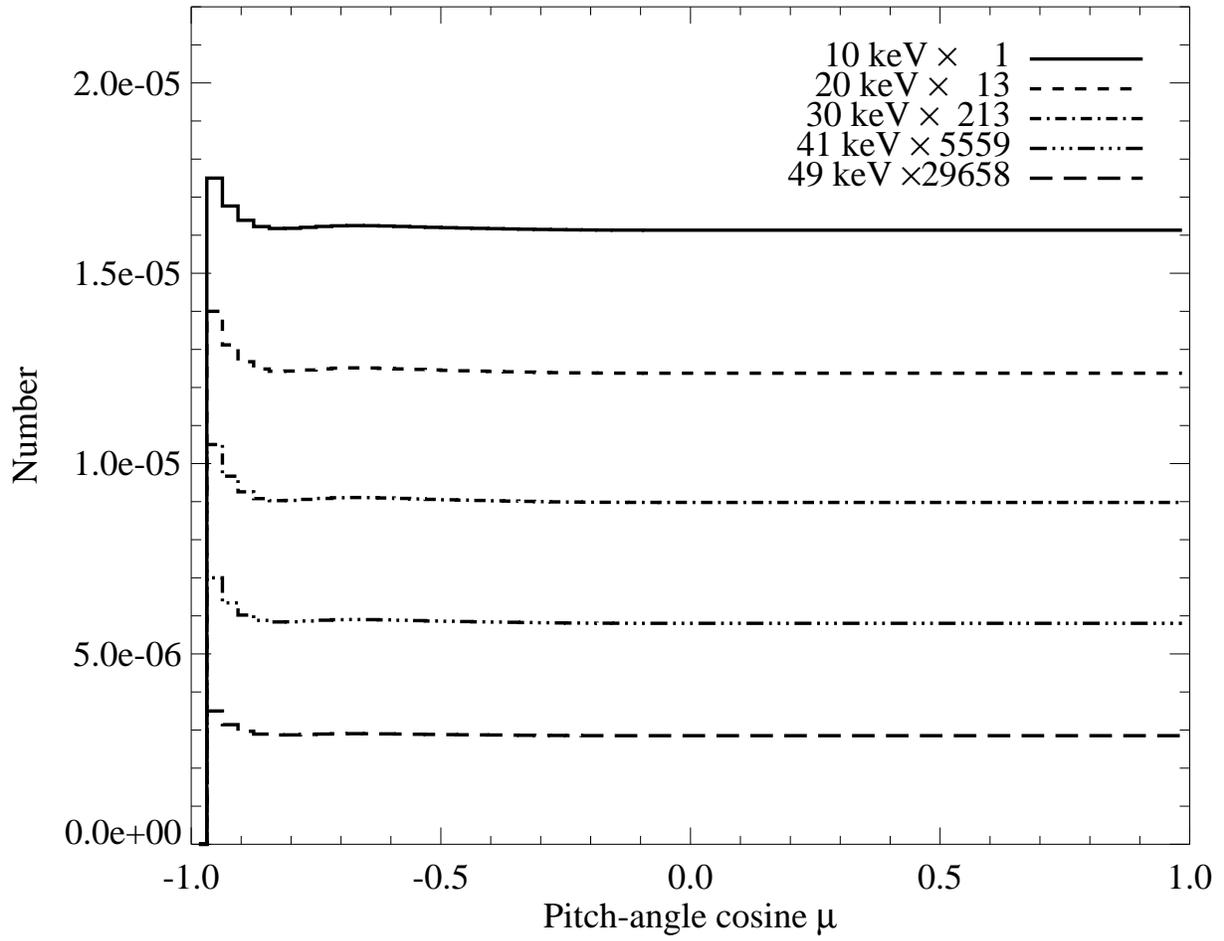}
\caption{Electron pitch-angle distribution calculated from the analytic solution of eq. (\ref{eq:19}). The format of this Figure is same as of Figure \ref{fig:psd_sp}.}
\label{fig:psd_sp_ana}
\end{figure}

Using equations (\ref{eq6}) - (\ref{eq:2}), we trace the position of electrons backward in time. Electrons are set at $(x,z) = (0.14,0.87)$ at $t = 5$ s, same as employed in Figure \ref{fig:psd_sp}. Determining the mirror points from this test particle simulation, we obtain $\tan \theta$ as a function of the electron initial pitch angle in Figure \ref{fig:tant}. The velocity of the mirror point certainly depends on the electron initial pitch angle.
Using this result, we can calculate the analytic solution of the electron distribution (eq. (\ref{eq:19})), which is presented in Figure \ref{fig:psd_sp_ana}. The solution shows an enhancement of the number of electrons with $\mu \sim -0.95$, similar to the simulation (Fig. \ref{fig:psd_sp}). This confirms that the simulation result is certainly interpreted as the above process. 
% On the other hand, the analytic solution does not show a decrease with $-0.8 \lsim \mu < 0$ seen in the simulation. 
% We consider that this is due to the propagation effect along the field line that is not treated in the analytic approach.

\subsection{Electron Escape from the Sun}\label{sec:electron-escape}
\begin{figure}[htbp]
\centering
\epsscale{1.0}
\plotone{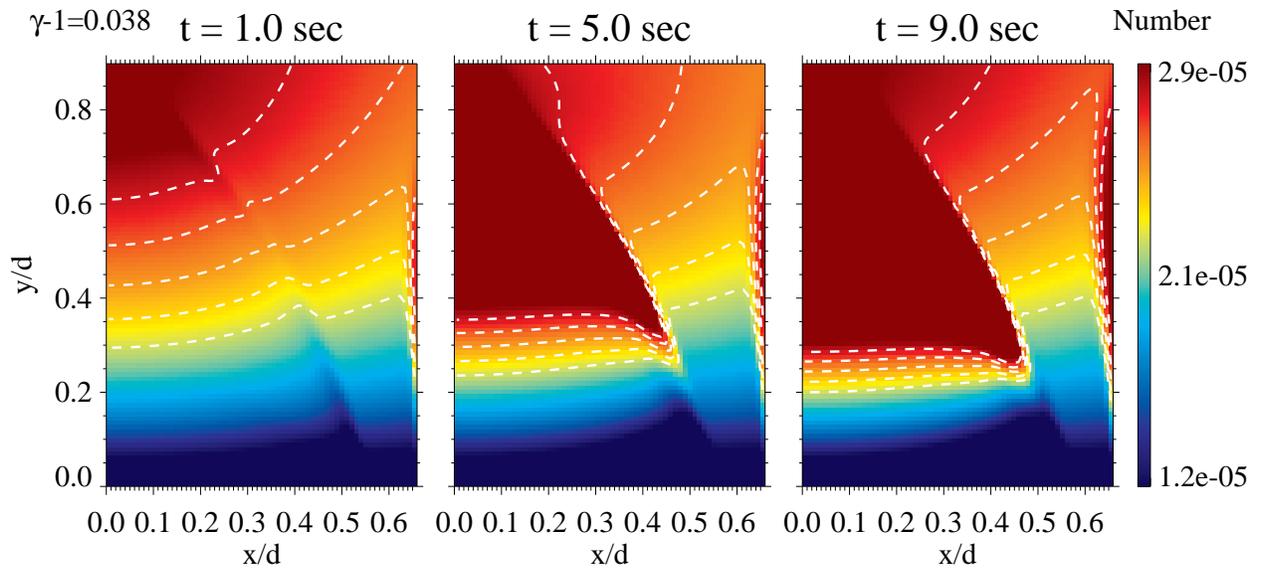}
\caption{Spatial distribution of the number of 20 keV antisunward electrons at $t=1$ s (left), 5 s (center), and 9 s (right). The color scale is set so as to emphasize the gradient in open field lines (right-half area). The white dashed lines show the contours of the number of electrons, with levels of 75\%, 80\%, 85\%, 90\% and 95\% of the maximum value.}
\label{fig:open_tim}
\end{figure}

The accelerated electrons due to the inertia drift propagate antisunward along open field lines, reach the upper boundary, and finally escape from the simulation domain. 
% It is important for the test through the comparison with observations to consider the number of escaping electrons and their time evolution. 
It is important to consider the number of these escaping electrons and their time evolution. 
Figure \ref{fig:open_tim} shows the time evolution of the spatial distribution of the number of 20 keV antisunward electrons. 
This figure emphasizes the gradient in open field lines (right-half area), to focus on the escaping electrons.

During the rising phase ($t = $ 1 - 5 s) the number of antisunward electrons decreases (identified as upward-shifting contours), meaning that the number of escaping electrons increases due to the growth of the electric field.
% Since escaping electrons are generated through the collision with the magnetic mirror points moving with $v_{\rm mirror} \propto v_E \propto E$, the acceleration rate is the highest when the electric field is the strongest at $t = t_p = 5$ s. 
The decrease of the electron number is more significant at the higher area, because the electric field is a monotonically increasing function of $z$. 
During the declining phase ($t = $ 5 - 9 s) the number of electrons recovers (identified as downward-shifting contours), because the electric field is reduced while constant electron fluxes are continuously injected from the upper boundary.
% The distribution is very similar between at $t = $ 1 and 9 s, at which the strength of the electric field is close.
The electron distribution at $t = $ 9 s is almost same as at $t = $ 1 s, because the electric field distribution is close.

\begin{figure}[htbp]
\centering
\epsscale{1.0}
\plotone{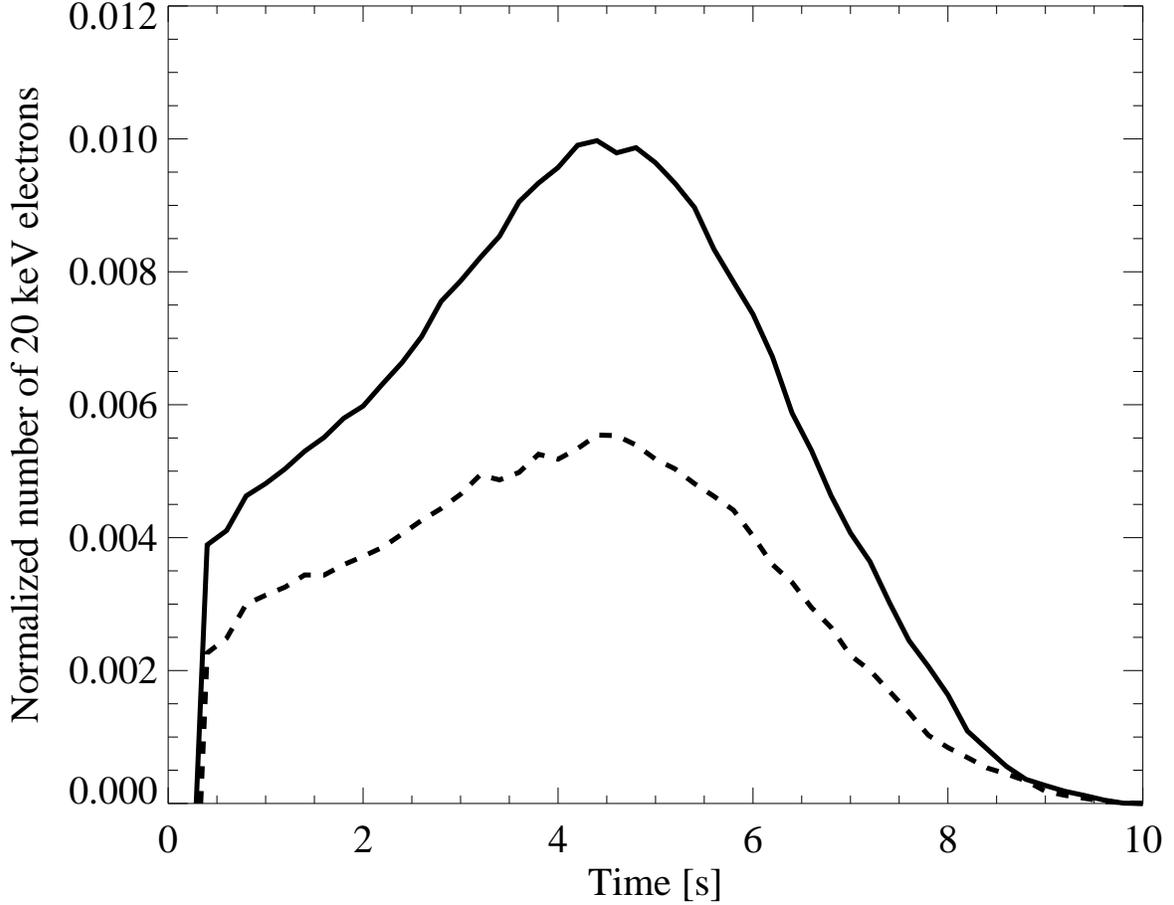}
\caption{Time profile of the number of 20 keV escaping electrons (eq. (\ref{eq:24})). The number is normalized to the initial distribution. The solid and dashed lines show the calculation results with $v_p = 60 \; {\rm and} \; 30 \; {\rm km \; s^{-1}}$, respectively. Just after the start of the simulation, electrons initially in the loss cone are lost at the footpoint and the number of electrons drastically decreases, because the initial distribution is not a steady state solution in the model. This appears as the discontinuous jumps at $t = 0.5$ s.}
\label{fig:escape_profile}
\end{figure}

The inertia drift acceleration takes place in a whole area of open field lines, especially near the magnetic separatrix (e.g., Fig. \ref{fig:psd_sp}).
We estimate the number of escaping electrons as
\begin{eqnarray}
f_{\rm escape}(\gamma,t) = -\int \!\!\! \int dxdz \left[\frac{1}{A(t)}\inty{-1}{0}f(t)d\mu - \frac{1}{A(10 \; {\rm s})}\inty{-1}{0}f(10 \; {\rm s})d\mu \right],\label{eq:24}
\end{eqnarray}
where $A(t)$ is the time-varying area of open field lines in the simulation domain, and the integration over the configuration space is implemented only in the area of open field lines. Here we use $f(10 \; {\rm s})$ as a reference, not the initial distribution $f(0 \; {\rm s})$, because the initial distribution is not a steady state solution in our model (see \S~\ref{sec:init-bound-cond}).
% The electric field is very weak and hence the distribution is very close to the steady state at $t = 10$ s. 
Figure \ref{fig:escape_profile} shows the time profile of the number of 20 keV escaping electrons. 
The solid line is obtained from the usual simulation result. The dashed line is from the result with the apparent velocity of the footpoint $v_p = 30 \; {\rm km \; s^{-1}}$, which is a half relative to the usual.

The time profile of the escaping electrons is similar to that of the apparent velocity of the footpoint, and thus the electric field (eqs. (\ref{eq:11}) and (\ref{eq:12})). This is understood as follows. A travel time of the escaping electrons in open field lines  is $\lsim 1$ s, which is much faster than the time scale of the temporal change of the electric field.
The accelerated electrons instantaneously reach the upper boundary and escape without being influenced by the temporal change of the electric field. 
Therefore the escaping electrons should directly reflect the instantaneous configuration of the electric field. 

The maximum number of the 20 keV escaping electrons is $\sim 1 \%$ of the initial. 
% We confirm that the ratio between the number of the escaping and initial electrons is $\sim 1 \%$ at a maximum within 10 - 50 keV. 
We confirm the similar percentage at maximum in 10 - 50 keV.
For electrons below 10 keV, the number is slightly small, $\sim 0.3 \%$.
Comparing the solid and dashed lines, it is found that the number of the escaping electrons is proportional to the electric field strength. 
As discussed, the rate of the inertia drift acceleration is simply proportional to the electric field ($v_{\rm mirror} \propto v_{E} \propto E$) so that it is expected that the number of escaping electrons correlates with the electric field. 
% The simulation results are in good agreement with the interpretation.

% We have to note discontinuous jumps just after $t = 0$ in Figure \ref{fig:escape_profile}.
% The initial uniform and isotropic condition is not a steady state solution in the model (\S \ref{sec:init-bound-cond}). Just after the start of the simulation, electrons initially in the loss cone are lost at the footpoint and the number of electrons drastically decreases. This decrease appears as the discontinuous jumps.

\section{Discussion}\label{sec:discussion}
The drift-kinetic simulation has identified two dominant mechanisms of electron acceleration in solar flares. One is the betatron acceleration at the top of closed loops, which enhances the electron velocity perpendicular to the magnetic field line. The other is the inertia drift acceleration in open magnetic field lines, which produces antisunward electrons. The resulting velocity space distribution significantly deviates from an isotropic distribution.

% Based on the drift-kinetic theory, we have developed a comprehensive model for particle acceleration and transport in solar flares. Using this model, we have investigated the time evolution of the electron distribution in electromagnetic fields of a flare. We have found from the simulation that there are two dominant mechanisms of electron acceleration. 
% The betatron acceleration, which generates higher energy electrons perpendicular to the magnetic field line, results in more confinement of them at the top of the loop. 

The betatron acceleration yields more confinement of higher energy electrons at the loop top, { as previously mentioned in the collapsing trap model \citep[][]{1997ApJ...485..859S,2004A&A...419.1159K,2006ApJ...647.1472K} }.
They can be a candidate for radiating loop-top nonthermal emissions such as ``above-the-loop-top'' HXR sources \citep{1994Natur.371..495M,1995PASJ...47..677M} and microwaves \citep[e.g.,][]{2001ApJ...557..880K,2002ApJ...576..505W,2002ApJ...580L.185M,2009ApJ...696..136H}. \cite{2008ApJ...673..598M} have found that the distribution of electrons is enhanced perpendicular to the field line when they are injected into a loop, from both observations of the 2003 May 29 flare and a numerical modeling of the electron transport with the {\FP} equation. 
This result can be interpreted that the betatron acceleration is the most dominant mechanism for the electron energization in the flare. 

The betatron acceleration at the loop top is quite reasonable to radiate loop-top nonthermal emissions. However, it is widely known that the most of HXR emissions are from footpoints of the loop. This indicates a large amount of electrons precipitating there. In the model almost all of electrons are trapped in the loop. One of the reasons for this discrepancy is that the model has not taken into account the pitch-angle scattering.
 The pitch-angle scattering leads electrons into the loss cone, and significantly increases the rate of the precipitation.
Many authors have argued the effect of the pitch-angle scattering of electrons by the Coulomb collisions with ambient plasma \citep{1976MNRAS.176...15M,1998ApJ...502..468A,1999ApJ...517..977A,2008ApJ...673..598M}. Another agency for the scattering is plasma waves. 
In particular, we suggest whistler waves as a possible candidate, because the perpendicular temperature anisotropy at the loop top in the simulation (Fig. \ref{fig:temp_dist}) is unstable for the whistler wave growth. 
The enhancements of the pitch-angle scattering by the interaction with whistler waves might efficiently work on the precipitation and subsequent HXR emissions at footpoints, similar to the precipitation of radiation belt electrons and diffuse auroral emissions at the Earth. We plan further simulations with including the pitch-angle scattering as diffusion terms in equation (\ref{eq:3}).

The number of escaping electrons produced by the inertia drift acceleration has been estimated up to $\sim 1$ \% of the background (Fig. \ref{fig:escape_profile}). The escaping electrons have been observed in association with flares, and their number is estimated to $\sim 0.1 - 1$ \% of the observed HXR-emitting electrons precipitating into the chromosphere \citep[e.g.,][]{1974SSRv...16..189L,2007ApJ...663L.109K}. 
The observed number of escaping electron is expected to be less than $\sim 1$ \% of the background, because the observed number of HXR-emitting electrons should be less than the background.
Therefore our estimation from the simulation can account for observations.
% Our estimation from the simulation is enough for observations, because the number of HXR-emitting electrons should not exceed the background. 
The inertia drift acceleration takes place in a whole area of open field lines and contributes this high efficiency, as long as curved magnetic field lines keeps moving.
% It does not necessarily require complex phenomena such as magnetic reconnection.
% It should be noted that a complex phenomenon such as magnetic reconnection is not necessary for the acceleration.
{It is noted that the observation can be understood without a complex phenomenon such as magnetic reconnection for the acceleration.}

Because of its ubiquity, we suggest the inertia drift acceleration as a possible mechanism for producing escaping electrons that are not always associated with flares \cite[e.g.,][]{1980ApJ...236L..97P,2003GeoRL..30m..30G,2007ApJ...657..567M,2010ApJ...708L..95E}.
 \cite{2007ApJ...657..567M} have reported the so-called ``micro type-III'' radio bursts. Micro type-III radio bursts, characterized by short-lived, continuous, and weak emissions, are thought to be an evidence of the ubiquitous electron acceleration in the solar corona. They have found that the micro type-III radio bursts are observed when the active regions bordering on coronal holes appear. It may suggest that parent electrons are accelerated near the boundary and then escape into interplanetary space. We have shown that antisunward electrons are produced especially near the magnetic separatrix, which qualitatively supports the observational results of \cite{2007ApJ...657..567M}.

{ In \S~\ref{sec:electron-escape} we have discussed that the time profile of the escaping electrons by the inertia drift acceleration reflects that of the electric field, because the travel time of the electrons is faster than the time scale of the temporal change of the electric field. As long as satisfying the condition, it is expected that the time scale of the escaping electrons becomes shorter following more rapid change of the electric field than we have employed ($\tau = 1.77 \; {\rm s}$). It may explain type-III radio bursts that have an elemental time scale shorter than $\sim 1$ s, while other mechanisms may also contribute to the acceleration. The escaping electrons may further be accelerated at a higher altitude, for example, the vicinity of the reconnection region and the possible turbulent region. The inertia drift acceleration can contribute to supply electrons to the higher corona and the interplanetary space where the secondary acceleration may occur.}

\cite{2009ApJ...697..843M} have reported that the electron distribution in the loop is enhanced parallel to the field line in the 2006 December 13 flare. They have argued that the distribution could be formed, if the inertia drift acceleration efficiently takes place. Such a distribution has not been formed in our model. In the drift-kinetic model, the betatron and inertia drift accelerations take place simultaneously; the former enhances the velocity distribution in the perpendicular direction while the latter in the parallel. 
The resulting distribution is determined by combination of these accelerations. 
We have found in \S \ref{sec:electr-distr-clos} that the rate of the betatron acceleration $\sim 6$ is more efficient than that of the inertia drift $\sim 1.3$ in the closed loop of the model magnetic field.
However, it may be expected that the inertia drift acceleration overcomes the betatron in different field configurations, and it is possible that the resulting electron distribution is enhanced rather parallel to the field line. 

\begin{figure}[htbp]
\centering
\epsscale{1.0}
\plotone{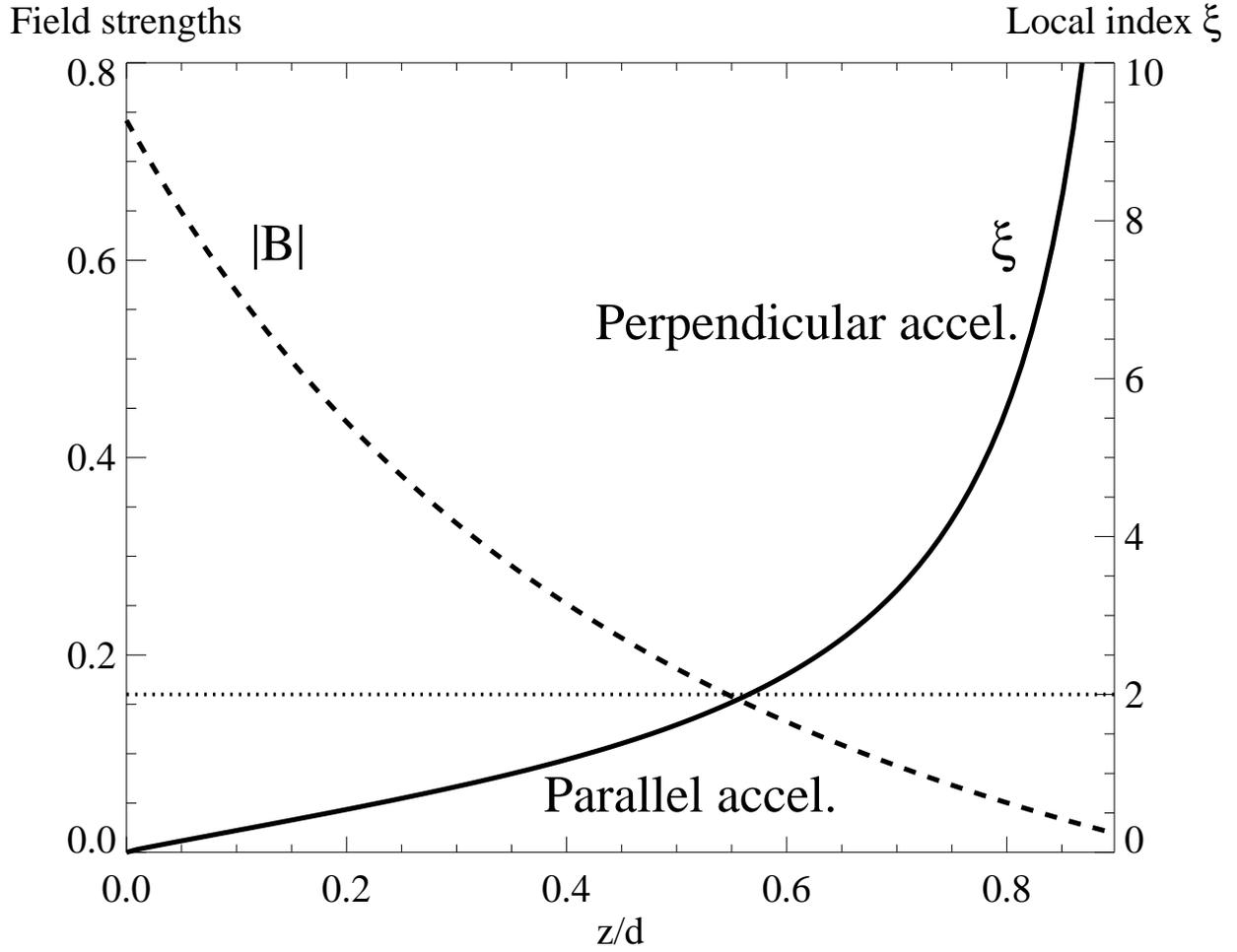}
\caption{Strength of the model magnetic field (dashed line, eqs. (\ref{eq:5}) - (\ref{eq:8})), and its local index $\xi$ (solid line) at the apex of loops ($x=0$) at $t=5$ s, as a function of the altitude. It is expected from equation (\ref{eq:29}) that particles are accelerated more perpendicular (parallel) to the magnetic field line at which $\xi$ is larger (smaller) than 2 (denoted as a dotted line).}
\label{fig:b_height}
\end{figure}

Let us simply consider the relationship between the magnetic field configuration of the loop and the distribution of accelerated particles.
Suppose that the magnetic field strength at the loop top is proportional to its altitude, $B \propto z^{-\xi}$. When the magnetic field is compressed by a factor of $R_c$, the altitude decreases by a factor of $R_c^{1/\xi}$. We assume that the travel length of particles along the field line $L$ is proportional to the altitude. From equations (\ref{eq:25}) and (\ref{eq:27}), we describe the increase of the perpendicular and parallel energies through the compression as
\begin{eqnarray}
\left\{
\begin{array}{l}
 \epsilon_{\perp}/\epsilon_{\perp,0} = R_c,\\
 \epsilon_{\parallel}/\epsilon_{\parallel,0} = R_c^{2/\xi}.\\
\end{array}
\right.\label{eq:29}
\end{eqnarray}
It is found that particles are accelerated perpendicular rather than parallel to the magnetic field line when $\xi > 2$.
Figure \ref{fig:b_height} shows the local index $\xi$ of the model magnetic field at the apex of loops ($x=0$). The index is much larger than 2 at the higher altitude where electrons are more accelerated by stronger electric fields (Fig. \ref{fig:xzimgs}). This is consistent with the perpendicular enhancement of the electron distribution function around the loop top, shown in Figure \ref{fig:psd_lt}.
Meanwhile, we propose from equation (\ref{eq:29}) that it may be possible to enhance the particle distribution parallel to the field line when the loop configuration meets $0 < \xi < 2$.
To verify this hypothesis, we should perform simulations with various field models, e.g., MHD simulation results \citep{2006ApJ...647.1472K}, and realistic fields extrapolated from {\Hinode} observations \citep{2008ASPC..397..110I,2008AGUFMSH52A..06K}. 
Toward the understanding of particle acceleration in solar flares through both the observations and simulations, it is critically important to study the relationship between a variety of magnetic field configurations and the resulting particle phase space distribution.

% Finally, we note that it is difficult to discuss the high energy distribution and the long-term evolution in the current simulation code, because of an unsatisfactory numerical accuracy. 
% To overcome these difficulties, we develope a new method to numerically solve the advection equation, especially Vlasov equation \citep{MMA}. The application of the method to this model is also under development, which will be reported in future.

{ Due to the numerical limitation, we have performed the simulation with a relatively high initial temperature 3 keV. Although the acceleration efficiency is dependent on the temperature, it does not mean that the pre-acceleration is necessary for the proposed acceleration mechanisms. In fact, we have also performed the simulation with an initial temperature 1.5 keV, and obtain similar results. This is to be expected, because the adiabatic betatron acceleration takes place similarly regardless of the particle energy. For the inertia drift acceleration, the particle velocity should be faster than the mirror point velocity to be reflected. Electrons can easily meet this condition. Therefore the acceleration takes place even for electrons with the temperature lower than we have employed in this study.}
 
% { Let us discuss the effect of the three-dimensionality of ambient fields, for example, shear motions and/or guiding magnetic fields $(B_y)$. In the three-dimensional system, electric fields $(-\vect{v} \times \vect{B})$ may have components in the two-dimensional plane ($E_x$ and $E_z$), which generate the voltage drop along the magnetic field line. Because of their high mobility, electrons will move against the drop. As a result, their distribution will be significantly modified from the two-dimensional system, particularly in the parallel direction. For the electron acceleration in the three-dimensional system, we have to consider the parallel electric field and the electrostatic potential that are ignored in the current two-dimensional model.}

{ Let us discuss the effect of the three-dimensionality of ambient fields, for example, the shear motion and/or guiding magnetic fields. In the three-dimensional system, the fields may evolve so as to generate field-aligned currents. The career of the current should be electrons, because of their high mobility. If electrons are inhibited to move along the field line by such as the magnetic mirror force and the pressure gradient (acting as a resistivity), parallel electric fields will be generated and accelerate electrons, to carry the current required from the field evolution \citep[e.g.,][]{1995PASJ...47..691T}. As a result, their distribution will be significantly modified from the two-dimensional system, particularly in the parallel direction. The concept of the current-voltage relation has been utilized especially for the auroral particle acceleration along the magnetic field line at the Earth \citep{1973P&SS...21..741K}. For the electron acceleration in the three-dimensional system, it may be essential to consider the parallel electric field and the electrostatic potential.}

\section{Conclusion}\label{sec:conclusion-1}
Based on the drift-kinetic theory, we have developed a comprehensive model for particle acceleration and transport in solar flares. Using this model, we have simulated the time evolution of the electron distribution in a flaring region. There are two dominant mechanisms of electron acceleration. 
The betatron acceleration takes place at the top of closed loops. It can be a generation mechanism of electrons that radiate loop-top nonthermal emissions. The phase space distribution of accelerated particles in the loop strongly depends on the magnetic field configuration.
The inertia drift acceleration also takes place in open magnetic field lines, producing escaping electrons from the Sun. The number of escaping electrons estimated from the simulation can account for the observed number of flare-associated escaping electrons. 
The inertia drift acceleration is caused by a motion of curved field lines, which is not necessarily driven by magnetic reconnection.
In this sense, we propose the acceleration in this study as a mechanism for producing escaping electrons that are not always associated with flares.

\begin{acknowledgements}
We thank to K. Kusano, T. Yokoyama, T. Amano, A. Morioka, and M. J. Aschwanden for fruitful discussions and comments. We also thank to anonymous referee for valuable comments to improve our manuscript.
T. M. is supported by the Grant-in-Aid for Young Scientists (B) \#21740135 and partly by the Grant-in-Aid for Creative Scientific Research of MEXT/Japan, the Basic Study of Space Weather Predication. This work is a part of the GEMSIS (Geospace Environment Modeling System for Integrated Studies) project of Solar-Terrestrial Environment Laboratory, Nagoya University.
\end{acknowledgements}

%Bibliography                                                                  
% \bibliographystyle{apj}                                                       
% \bibliography{ref}   

% Figures

% \begin{figure}[htbp]
% \centering
% \epsscale{1.0}
% \plotone{./xzimg_20keV.eps}
% \caption{Spatial distribution of the number of 20 keV omnidirectional electrons at $t = 5 \; {\rm s}$  (left) and $10 \; {\rm s}$ (right). The white lines are the magnetic field lines, and the dashed lines are the magnetic separatrix.}
% \label{fig:xzimg_20kev}
% \end{figure}

% \begin{figure}[htbp]
% \centering
% \epsscale{1.0}
% \plotone{./xzimg_49keV.eps}
% \caption{The same as Figure \ref{fig:xzimg_20kev}, but for 50 keV electrons.}
% \label{fig:xzimg_49kev}
% \end{figure}

\end{document}